\documentclass[aps,prx,showpacs,twocolumn,amsmath,amssymb,nofootinbib]{revtex4-1}
\usepackage{graphicx}
\usepackage{dcolumn}
\usepackage{bm}
\usepackage{amsmath}
\usepackage{amsfonts}
\usepackage{amssymb}
\usepackage{multirow}

\usepackage{amsthm}
\usepackage{pinlabel}
\usepackage{natbib}

\usepackage{epstopdf}
\usepackage[abs]{overpic}
\usepackage[dvipsnames]{xcolor}

\usepackage{tikz}
\usetikzlibrary{arrows.meta}
\usetikzlibrary{decorations.pathmorphing}
\usetikzlibrary{decorations.markings}

\tikzset{
	particle/.style={thick,draw=blue, postaction={decorate},
		decoration={markings,mark=at position .5 with {\arrow[blue]{triangle 45}}}},
	gluon/.style={decorate, draw=black,
		decoration={coil,aspect=0}}
}

\usepackage[dvipsnames]{xcolor}

\usepackage{braket}

\definecolor{specialgray}{HTML}{505050}
\definecolor{col10K}{HTML}{FFA000}
\definecolor{col300K}{HTML}{924FA4}
\definecolor{colMu}{HTML}{5278BD}
\definecolor{colMuI}{HTML}{924FA4}

\definecolor{specialgray}{HTML}{505050}
\definecolor{col10K}{HTML}{FFA000}
\definecolor{col300K}{HTML}{924FA4}
\definecolor{colMu}{HTML}{5278BD}
\definecolor{colMuI}{HTML}{924FA4}
\definecolor{newred}{HTML}{D53E4F}
\definecolor{newblue}{HTML}{5278BD}
\definecolor{newcyan}{HTML}{1EA0A0}
\definecolor{newgreen}{HTML}{5CB14E}
\definecolor{newpurple}{HTML}{924FA4}
\definecolor{newyellow}{HTML}{D1C72E}
\definecolor{neworange}{HTML}{D6923C}

\begin{document}

\title{Influence of phonon renormalization in Eliashberg theory for superconductivity in\\
 2D and 3D systems}
\author{Fabian Schrodi}\email{fabian.schrodi@physics.uu.se}
\author{Alex Aperis}\email{alex.aperis@physics.uu.se}
\author{Peter M. Oppeneer}\email{peter.oppeneer@physics.uu.se}
\affiliation{Department of Physics and Astronomy, Uppsala University, P.\ O.\ Box 516, SE-75120 Uppsala, Sweden}

\vskip 0.4cm
\date{\today}

\begin{abstract}
	\noindent 
	
	Eliashberg's foundational theory of superconductivity is based on the application of Migdal's approximation, which states that vertex corrections to first order electron-phonon scattering are negligible if the ratio between phonon and electron energy scales is small. The resulting theory incorporates the first Feynman diagrams for electron and phonon self-energies. However, the latter is most commonly neglected in numerical analyses. Here we provide an extensive study of full-bandwidth Eliashberg theory in two and three dimensions, where we include the full back reaction of electrons onto the phonon spectrum. We unravel the complex interplay between nesting properties, Fermi surface density of states, renormalized electron-phonon coupling, phonon softening, and superconductivity. We propose furthermore a scaling law for the maximally possible critical temperature $T_c^{\textrm{max}}\propto\lambda (\Omega ) \sqrt{\Omega_0^2-\Omega^2}$ in 2D and 3D systems, which embodies both the renormalized electron-phonon coupling strength $\lambda(\Omega)$ and softened phonon spectrum $\Omega$. Also, we analyze for which electronic structure properties  a maximal $T_c$ enhancement can be achieved.
\end{abstract}

\maketitle

\section{Introduction}

The current state-of-the-art description of superconductors is Eliashberg theory\,\cite{Eliashberg1960}, which is especially applied in cases where the more simplified BCS (Bardeen-Cooper-Schrieffer) treatment\,\cite{Bardeen1957} cannot capture the main characteristics of a given system. One of the key aspects to the success of Eliashberg theory is the applicability of Migdal's approximation\,\cite{Migdal1958}, which states that higher order Feynman diagrams for electron-phonon scattering can be neglected if the ratio of phonon to electron energy scale is a small number. In such a case it is sufficient to only consider all first-order Feynman diagrams for the electron and phonon self-energy, although the latter is neglected in most cases. Such a neglect may be motivated by the drive for making an extremely complicated problem easier. Interestingly though, in the original works by Migdal\,\cite{Migdal1958} and Eliashberg\,\cite{Eliashberg1960} the phonon self-energy was included in the calculation in an approximative way. However, for a quantitative analysis it is a generally accepted procedure to neglect the back reaction of electrons onto the phonon spectrum.

In available literature the phonon renormalization is most commonly considered only when checking the validity of Eliashberg theory calculations\,\cite{Marsiglio1990,Scalettar1989,Marsiglio2020,Chubukov2020}. The numerical results are then benchmarked against outcomes of Quantum Monte Carlo (QMC) \,\cite{Esterlis2018,Esterlis2018_2} or Dynamical Mean-Field Theory (DMFT) simulations\,\cite{Bauer2011}. Comparing conclusions from various authors does not necessarily lead to a completely coherent picture concerning the validity of Eliashberg theory, but there is consensus about the existence of a maximal electron-phonon coupling strength marking the border of applicability. Characteristics of the superconducting state in such studies are found only by extrapolation from normal state properties, and, in addition, the QMC calculations are performed on relatively small lattices due to the huge complexity of the problem. The most commonly studied system in these works is the 2D Holstein model\,\cite{Alexandrov2001,Meyer2002,Perroni2005,Costa2018,Esterlis2018,Dee2019,Dee2020}, often with additional constraints such as half-filling, while 3D systems are rarely considered in numerical calculations, presumably due to the large computational complexity.

Another way of checking the validity of the commonly employed Migdal approximation, and hence of the resulting Eliashberg theory, is to compute vertex corrections corresponding to additional Feynman diagrams. This has been attempted in various works under different kinds of approximation\,\cite{Kostur1993,Gunnarsson1994,Nicol1994,Miller1998,Hague2003,Pietronero1992_2,Benedetti1994,Paci2001,Boeri2003,Cappelluti2003}. The first vertex-corrected self-consistent Eliashberg theory without further simplifications has been recently proposed by the current authors\,\cite{Schrodi2020_2}. All these works have in common that one or more additional Feynman diagrams to the electron-phonon interaction are studied and compared to the commonly employed adiabatic formalism, that does not include a finite phonon self-energy.

The aim of our current work is to give a comprehensive overview of the influence of phonon renormalization occurring within adiabatic Eliashberg theory when the first-order Feynman diagram for the phonon self-energy is taken self-consistently into account, rather than commenting much on its validity with respect to other theories. Our efficient implementation allows not only to access sufficiently low temperatures to study the superconducting state without relying on extrapolation of normal state properties, but also opens the discussion of 3D systems, which has so far been elusive in available literature. Our theory is based on a Holstein-like Hamiltonian, containing a nearest neighbor tight-binding model for the electron energies, an isotropic Einstein phonon mode and isotropic electron-phonon scattering elements, controlling the coupling strength in the system. We explore here in detail the interplay of phonon softening, renormalized coupling strength, nesting properties, superconducting energy gap, and, eventually, the maximally possible transition temperature $T_c$. 

Our theory takes the bare phonon frequency $\Omega_0$, coupling strength $\lambda_0$ and electron energies $\xi_{\mathbf{k}}$ as inputs. By varying these quantities we pinpoint the renormalized coupling strength and phonon softening (momentum dependent decrease in magnitude) as key ingredients to how our self-consistent results change. Other crucially important aspects for the interacting state are Fermi surface (FS) nesting properties of $\xi_{\mathbf{k}}$ and the density of states (DOS) at the Fermi level. There exists a critical value $\lambda_0^{\star}$ of the bare input coupling, at which the phonon energies become negative, indicating a lattice instability. Another important value of $\lambda_0$ marks the onset of superconductivity, $\lambda_0^{\Delta}$. In 2D we find an enhancement of both $\lambda_0^{\star}$ and $\lambda_0^{\Delta}$ with increased shallowness of $\xi_{\mathbf{k}}$, which goes along with less coherent nesting properties and a decrease in FS DOS. Notably, we find that our model systems exhibit maximal superconducting transition temperatures $T_c$ for an intermediate system that is not very shallow, but also not ideally nested. For 3D systems the significance of FS nesting is weakened, such that trends in $\lambda_0^{\star}$, $\lambda_0^{\Delta}$ and maximum $T_c$ are mainly dictated by the FS DOS. An exception to these tendencies is our most shallow $\xi_{\mathbf{k}}$ considered in 3D, because in this particular system nesting is exceptionally coherent due to the special role taken by the $\Gamma$ point of the Brillouin zone (BZ).

From here we proceed as follows: In the next Section \ref{scTheory} we introduce the formalism and mathematical steps needed for deriving a self-consistent and full bandwidth Eliashberg theory, that includes all first-order processes for electron and phonon self-energies. We provide some benchmark checks of our implementation in Appendix \ref{scAppBench}. We continue by introducing electron dispersions in two and three dimensions, see Section \ref{scModel}. In particular, we define in both cases three different energies via varying the chemical potential, that differ in nesting and FS properties. In Section \ref{scSoftCoup} follows an exploration of phase space, spanned by the different input parameters to our theory. We discuss in detail the aspects of phonon softening, renormalization of electron-phonon coupling, and the onset of superconductivity, and try to unravel their complex interplay by deriving approximate relations between them. In the following Section \ref{scTc} the subject is a closer discussion of the superconducting energy gap as function of temperature, and consequently the transition temperatures, in 2D and 3D. Motivated by the precedent findings we propose a scaling law for the maximum transition temperature that models our numerical results to very good accuracy. The paper is concluded in Section \ref{scConclusion} with a brief discussion on related works, possible extensions to our theory, and potential future directions.

\section{Theory}\label{scTheory}

We consider a phonon mode with frequency $\Omega_0$ and electron-phonon coupling that is given by $g_{\mathbf{q}}$. Here the electronic energies are modeled by a single band dispersion $\xi_{\mathbf{k}}$, with $\mathbf{k}$ a BZ wave vector. By setting $\mathbf{q}=\mathbf{k}-\mathbf{k}'$ we can write the Hamiltonian as
\begin{align}
H =& \sum_{\mathbf{k}}\xi_{\mathbf{k}} \Psi^{\dagger}_{\mathbf{k}}\hat{\rho}_3\Psi_{\mathbf{k}} + \sum_{\mathbf{q}} \hbar\Omega_0\Big( b^{\dagger}_{\mathbf{q}}b_{\mathbf{q}} + \frac{1}{2} \Big) \nonumber\\
&+  \sum_{\mathbf{k},\mathbf{k}'} g_{\mathbf{k}-\mathbf{k}'}  u_{\mathbf{k}-\mathbf{k}'} \Psi^{\dagger}_{\mathbf{k}'}\hat{\rho}_3\Psi_{\mathbf{k}} ~.
\end{align}
Above we use $b^{\dagger}_{\mathbf{q}}$ and $b_{\mathbf{q}}$ as bosonic creation and annihilation operators, which determine the phonon displacement $u_{\mathbf{q}}=b_{\mathbf{q}}+b_{-\mathbf{q}}^{\dagger}$. The electronic creation and annihilation operators $c_{\mathbf{k},\sigma}^{\dagger}$ and $c_{\mathbf{k},\sigma}$ are part of the Nambu spinor $\Psi^{\dagger}_{\mathbf{k}}=\big(c^{\dagger}_{\mathbf{k},\uparrow},c_{-\mathbf{k},\downarrow}\big)$, where $\sigma\in\{\uparrow,\downarrow\}$ is a spin label.
The imaginary time $\tau$ dependent electron and phonon Green's functions are respectively given by
\begin{align}
\hat{G}_{\mathbf{k}}(\tau) &= -\langle\mathcal{T}_{\tau} \Psi_{\mathbf{k}}(\tau) \otimes \Psi^{\dagger}_{\mathbf{k}}(0)\rangle~, \\
D_{\mathbf{q}}(\tau) &= -\langle\mathcal{T}_{\tau} u_{\mathbf{q}}(\tau) u_{\mathbf{q}}(0)\rangle ~,
\end{align}
where $\mathcal{T}_{\tau}$ is the imaginary time ordering operator. Both propagators obey a Dyson equation  in Matsubara space, reading
\begin{align}
\hat{G}_{\mathbf{k},m} &= \hat{G}^0_{\mathbf{k},m} + \hat{G}^0_{\mathbf{k},m}  \hat{\Sigma}_{\mathbf{k},m} \hat{G}_{\mathbf{k},m} , \label{dysonG} \\
D_{\mathbf{q},l} &= D^0_{\mathbf{q},l} + D^0_{\mathbf{q},l} \Pi_{\mathbf{q},l} D_{\mathbf{q},l} ~, \label{dysonD}
\end{align}
where we make use of the notation $f(\mathbf{k},i\omega_m)=f_{\mathbf{k},m}$ with Matsubara frequencies $\omega_m=\pi T(2m+1)$ for fermions, and $g(\mathbf{q},iq_l)=g_{\mathbf{q},l}$ for bosons with $q_l=2\pi Tl$. The above Eqs.\,(\ref{dysonG}) and (\ref{dysonD}) are solved for the dressed propagators as functions of the respective self-energies $\hat{\Sigma}_{\mathbf{k},m}$ and $\Pi_{\mathbf{q},l}$:
\begin{align}
\hat{G}^{-1}_{\mathbf{k},m} &= [\hat{G}^0_{\mathbf{k},m}]^{-1} - \hat{\Sigma}_{\mathbf{k},m} , \label{invG} \\
D^{-1}_{\mathbf{q},l} &= [D^0_{\mathbf{q},l}]^{-1} - \Pi_{\mathbf{q},l} . \label{invD}
\end{align}

The non-interacting Green's functions are defined by
\begin{align}
[\hat{G}^0_{\mathbf{k},m}]^{-1} &= i\omega_m\hat{\rho}_0 - \xi_{\mathbf{k}}\hat{\rho}_3 , \label{bareG} \\
[D^0_{\mathbf{q},l}]^{-1}  &= \! \Big(\frac{1}{iq_l - \Omega_0} - \frac{1}{iq_l + \Omega_0}\Big)^{-1}\! = \!-\frac{1}{2\Omega_0}\big(q_l^2 + \Omega_0^2\big)  , \label{bareD}
\end{align}
where we use the Pauli matrices $\hat{\rho}_i$, $i\in\{0,1,2,3\}$. The electronic self-energy can be decomposed to the mass enhancement function $Z_{\mathbf{k},m}$, chemical potential renormalization $\Gamma_{\mathbf{k},m}$, and superconducting order parameter $\phi_{\mathbf{k},m}$:
\begin{align}
\hat{\Sigma}_{\mathbf{k},m} = i\omega_m \big(1-Z_{\mathbf{k},m}\big) \hat{\rho}_0 + \Gamma_{\mathbf{k},m}\hat{\rho}_3 + \phi_{\mathbf{k},m}\hat{\rho}_1 . \label{sigmaDef}
\end{align}
Together with Eq.\,(\ref{invG}) this leads to an inverse electron Green's function
\begin{align}
\hat{G}^{-1}_{\mathbf{k},m} = i\omega_mZ_{\mathbf{k},m}\hat{\rho}_0 - \big(\xi_{\mathbf{k}} + \Gamma_{\mathbf{k},m}\big)\hat{\rho}_3 - \phi_{\mathbf{k},m}\hat{\rho}_1 ~,
\end{align}
so that after matrix inversion we get
\begin{align}
\hat{G}_{\mathbf{k},m} &= \Big[ i\omega_mZ_{\mathbf{k},m}\hat{\rho}_0 + \big(\xi_{\mathbf{k}}+\Gamma_{\mathbf{k},m}\big)\hat{\rho}_3 +\phi_{\mathbf{k},m}\hat{\rho}_1 \Big] \Theta^{-1}_{\mathbf{k},m} , \label{elecGreen} \\
\Theta_{\mathbf{k},m} &= \big(i\omega_mZ_{\mathbf{k},m}\big)^2 - \big(\xi_{\mathbf{k}}+\Gamma_{\mathbf{k},m}\big)^2 - \phi^2_{\mathbf{k},m} ~. \label{theta}
\end{align}

\begin{figure}[h!]
	\centering
	\includegraphics[width=0.9\linewidth]{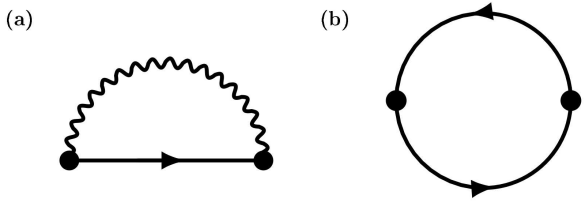}
	\caption{(a), (b) First order Feynman diagram for the electron and phonon self-energy, respectively.}
	\label{feynman}
\end{figure}

In this work we take into account all first order processes in both the electron and phonon self-energies. The corresponding Feynman diagrams are shown in Fig.\,\ref{feynman}(a) and (b). 
By exploiting momentum and energy conservation in the scattering processes we can translate the Feynman diagram for the electron self-energy into
\begin{align}
\hat{\Sigma}_{\mathbf{k},m} &= -T  \sum_{\mathbf{k}',m'} |g_{\mathbf{k}-\mathbf{k}'}|^2 D_{\mathbf{k}-\mathbf{k}',m-m'} \hat{\rho}_3 \hat{G}_{\mathbf{k}',m'} \hat{\rho}_3  , \label{sigma1}
\end{align}
and for the phonon self-energy we obtain
\begin{align}
\Pi_{\mathbf{q},l} &= T |g_{\mathbf{q}}|^2 \sum_{\mathbf{k},m} \mathrm{Tr} \left\{ \hat{\rho}_3 \hat{G}_{\mathbf{k},m} \hat{\rho}_3 \hat{G}_{\mathbf{k}+\mathbf{q},m+l} \right\}  . \label{pidef}
\end{align}

At this point it is convenient to define the electron-phonon interaction kernel
\begin{align}
V_{\mathbf{q},l}=-|g_{\mathbf{q}}|^2D_{\mathbf{q},l} ,\label{kern}
\end{align}
so that we can write Eq.\,(\ref{sigma1}) as
\begin{align}
\hat{\Sigma}_{\mathbf{k},m} &= T  \sum_{\mathbf{k}',m'}  V_{\mathbf{k}-\mathbf{k}',m-m'} \hat{\rho}_3 \hat{G}_{\mathbf{k}',m'} \hat{\rho}_3  . \label{sigma2}
\end{align}
Combining Eq.\,(\ref{sigma2}) with Eqs.\,(\ref{sigmaDef}) and (\ref{elecGreen}) leads to the Eliashberg equations
\begin{align}
Z_{\mathbf{k},m} &= 1 - \frac{T}{\omega_m} \sum_{\mathbf{k}',m'} V_{\mathbf{k}-\mathbf{k}',m-m'} \frac{\omega_{m'}Z_{\mathbf{k}',m'}}{\Theta_{\mathbf{k}',m'}} , \label{zz}\\
\Gamma_{\mathbf{k},m} &= T \sum_{\mathbf{k}',m'} V_{\mathbf{k}-\mathbf{k}',m-m'} \frac{\xi_{\mathbf{k}'}+\Gamma_{\mathbf{k}',m'}}{\Theta_{\mathbf{k}',m'}} , \\
\phi_{\mathbf{k},m} &= -T \sum_{\mathbf{k}',m'} V_{\mathbf{k}-\mathbf{k}',m-m'} \frac{\phi_{\mathbf{k}',m'}}{\Theta_{\mathbf{k}',m'}} ,\label{phi}
\end{align}
from which the superconducting gap function can be defined as $\Delta_{\mathbf{k},m}=\phi_{\mathbf{k},m}/Z_{\mathbf{k},m}$. Inserting Eq.\,(\ref{elecGreen}) into Eq.\,(\ref{pidef}) gives the phonon self-energy as function of $Z_{\mathbf{k},m}$, $\Gamma_{\mathbf{k},m}$ and $\phi_{\mathbf{k},m}$:
\begin{align}
\Pi_{\mathbf{q},l} &= -2T |g_{\mathbf{q}}|^2 \sum_{\mathbf{k},m} \Bigg(  \frac{\omega_{m}Z_{\mathbf{k},m}}{\Theta_{\mathbf{k},m}}\frac{\omega_{m+l}Z_{\mathbf{k}+\mathbf{q},m+l}}{\Theta_{\mathbf{k}+\mathbf{q},m+l}} \nonumber  \\
&- \frac{\xi_{\mathbf{k}}+\Gamma_{\mathbf{k},m}}{\Theta_{\mathbf{k},m}}\frac{\xi_{\mathbf{k}+\mathbf{q}}+\Gamma_{\mathbf{k}+\mathbf{q},m+l}}{\Theta_{\mathbf{k}+\mathbf{q},m+l}} + \frac{\phi_{\mathbf{k},m}}{\Theta_{\mathbf{k},m}}\frac{\phi_{\mathbf{k}+\mathbf{q},m+l}}{\Theta_{\mathbf{k}+\mathbf{q},m+l}}\Bigg) . \label{pifinal}
\end{align}

The above equations are solved self-consistently in an iterative manner: Assuming $Z_{\mathbf{k},m}$, $\Gamma_{\mathbf{k},m}$ and $\phi_{\mathbf{k},m}$ are known from a previous iteration, we first calculate the phonon self-energy via Eq.\,(\ref{pifinal}), by using also Eq.\,(\ref{theta}). With $\Pi_{\mathbf{q},l}$ at hand, and the bare phonon propagator in Eq.\,(\ref{bareD}), we calculate the phonon Green's function from Eq.\,(\ref{invD}), which then determines the electron-phonon interaction kernel Eq.\,(\ref{kern}). As final step we solve for the mass renormalization, the chemical potential renormalization and the gap function via Eqs.\,(\ref{zz}-\ref{phi}). This process is repeated until convergence is reached. From the results we can calculate the electron filling as
\begin{align}
n = 1 - 2T\sum_{\mathbf{k},m} \frac{\xi_{\mathbf{k}}+\Gamma_{\mathbf{k},m}}{\Theta_{\mathbf{k},m}}~. \label{n1}
\end{align}

The input to our theory is the electron dispersion $\xi_{\mathbf{k}}$, the phonon frequency $\Omega_0$ and the coupling $\lambda_0$, which can be expressed as
\begin{align}
\lambda_0 = \langle\langle \lambda_{\mathbf{k}-\mathbf{k}'} \rangle_{\mathbf{k}\in \mathrm{FS}}\rangle_{\mathbf{k}'\in\mathrm{FS}} = \frac{2N_0}{\Omega_0} \langle\langle  |g_{\mathbf{k}-\mathbf{k}'}|^2  \rangle_{\mathbf{k}\in \mathrm{FS}}\rangle_{\mathbf{k}'\in\mathrm{FS}} ~, \label{lambda0}
\end{align}
with $N_0$ the DOS at the FS. After solving self-consistently for the interacting state we have access to the renormalized coupling strength $\lambda_{\mathbf{q}}$ and frequencies $\Omega_{\mathbf{q}}$. These are respectively given as
\begin{align}
\lambda_{\mathbf{q}} &= - N_0 |g_{\mathbf{q}}|^2 D_{\mathbf{q},l=0} ~,\label{lambdatilde} \\
\Omega_{\mathbf{q}} &= \sqrt{\Omega_0^2 + 2\Omega_0 \Pi_{\mathbf{q},l=0}  } ~, \label{omegatilde}
\end{align}
and it is convenient to define $\lambda=\langle\langle \lambda_{\mathbf{k}-\mathbf{k}'} \rangle_{\mathbf{k}\in \mathrm{FS}}\rangle_{\mathbf{k}'\in\mathrm{FS}}$ as a measure of the total coupling strength in the system. A non-vanishing imaginary part of $\Omega_{\mathbf{q}}$ marks a lattice instability.

The theory presented here has been included in the Uppsala Superconductivity Code (UppSC)\,\cite{UppSC,Aperis2015,Bekaert2018,Schrodi2019,Schrodi2020_3,SchrodiMultiChan}, and we benchmark our implementation in the Appendix \ref{scAppBench} against existing literature
\cite{Esterlis2018}.

\section{Model systems}\label{scModel}

\begin{figure}[b!]
	\centering
	\includegraphics[width=1\linewidth]{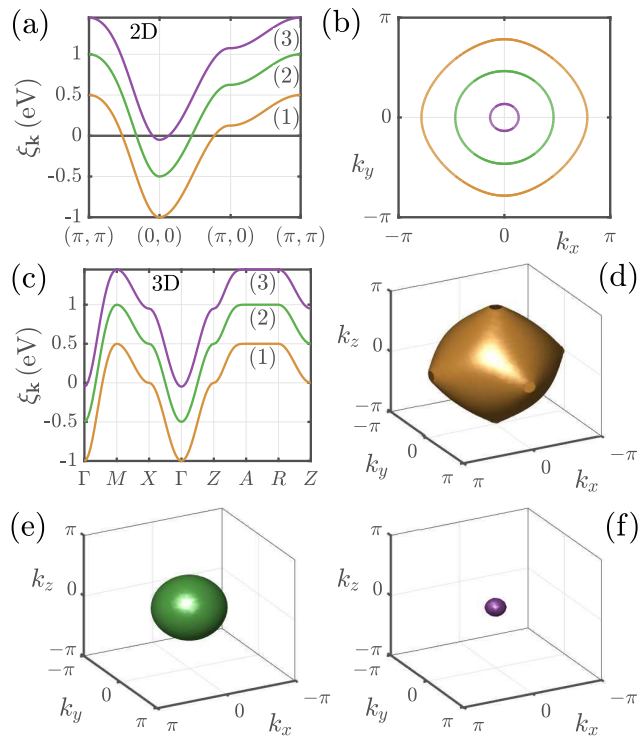}
	\caption{(a) Two dimensional electron energies along high symmetry lines of the Brillouin zone. The purple, green and orange curves are found from Eq.\,(\ref{xi2D}) by choosing $\mu=-887.5\,\mathrm{meV}$, $\mu=-437.5\,\mathrm{meV}$ and $\mu=62.5\,\mathrm{meV}$. (b) Fermi surfaces corresponding to panel (a), drawn in similar color code. (c) Three dimensional electron dispersion $\xi_{\mathbf{k}}$ as calculated from Eq.\,(\ref{xi3d}) for different chemical potentials (orange: $\mu=-0.125\,\mathrm{eV}$, green: $\mu=-0.625\,\mathrm{eV}$, purple: $\mu=-1.075\,\mathrm{eV}$), shown along high symmetry lines. (d)-(f) Three dimensional Fermi surfaces colored in correspondence to panel (c).}\label{energies}
\end{figure}

In this section we introduce two kinds of electron dispersions that are employed in the rest of our work. Starting in two spatial dimensions we define the electron energies as
\begin{align}
\xi_{\mathbf{k}} = -t^{(1)} \big[\cos(k_x) + \cos(k_y)\big] - t^{(2)} \cos(k_x)\cos(k_y) - \mu  ~, \label{xi2D}
\end{align}
where $t^{(1)}$ and $t^{(2)}$ are the nearest and next-nearest neighbor hopping energies of our tight-binding model, and $\mu$ is the chemical potential. Whenever referring to a 2D system, we choose $t^{(1)}=W/4$ and $t^{(2)}=t^{(1)}/2$, where $W=1.5\,\mathrm{eV}$ is the electronic bandwidth. We test three different examples for the choice of $\mu$, the resulting dispersions along high symmetry lines of the BZ are shown in Fig.\,\ref{energies}(a), while corresponding Fermi surfaces (same color code) are plotted in panel (b). The orange (1), green (2) and purple (3) curves are respectively found using $\mu=62.5\,\mathrm{meV}$, $\mu=-437.5\,\mathrm{meV}$ and $\mu=-887.5\,\mathrm{meV}$.

Turning now to the case of three spatial dimensions, we use the electron energies
\begin{align}
\xi_{\mathbf{k}} =& -t^{(1)} \sum_{i=x,y,z}\cos(k_i) \nonumber \\
&-t^{(2)} \sum_{i=x,y,z}\prod_{j=x,y,z; j\neq i} \cos(k_j) - \mu  ~. \label{xi3d}
\end{align}
Similarly as before we use an electronic bandwidth $W=1.5\,\mathrm{eV}$ and fix the hopping energies as $t^{(1)}=W/4$, $t^{(2)}=t^{(1)}/2$. We define three different dispersions that are shown in Fig.\,\ref{energies}(c) along high symmetry lines of the BZ. The orange (1), green (2) and purple (3) line represent the choices $\mu=-0.125\,\mathrm{eV}$, $\mu=-0.625\,\mathrm{eV}$ and $\mu=-1.075\,\mathrm{eV}$, respectively. The FS shown in Fig.\,\ref{energies}(d) is very large and resembles to a good approximation an ideal Fermi gas. As the opposite case, we have an extremely shallow energy dispersion reflected by the tiny FS in Fig.\,\ref{energies}(f). In Fig.\,\ref{energies}(e) we show an intermediate case. 

In the following we examine how the choice of electron dispersion affects our self-consistent results of the Eliashberg equations in two and three spatial dimensions. For simplification we consider here isotropic electron-phonon scattering, $g_{\mathbf{q}}=g_0$. As stated in Section\,\ref{scTheory}, the input to our Eliashberg equations is the bare electron-phonon coupling $\lambda_0$, the electron dispersion $\xi_{\mathbf{k}}$ and the bare phonon frequency $\Omega_0$. Since we want to study trends with respect to those input parameters we need to pay special attention to the electron filling that is calculated as function of the self-consistent results, see Eq.\,(\ref{n1}). To be able to compare outcomes for different $\Omega_0$, $\lambda_0$ and $T$ we need to ensure that $n$ stays constant, which we achieve by introducing an additional chemical potential shift $\delta\mu$, such that the input to the Eliashberg equations is $\xi_{\mathbf{k}}-\delta\mu$\,\cite{Schrodi2018}.

\section{Phonon softening, coupling and superconducting gap}\label{scSoftCoup}

\begin{figure}[h!]
	\centering
	\includegraphics[width=1\linewidth]{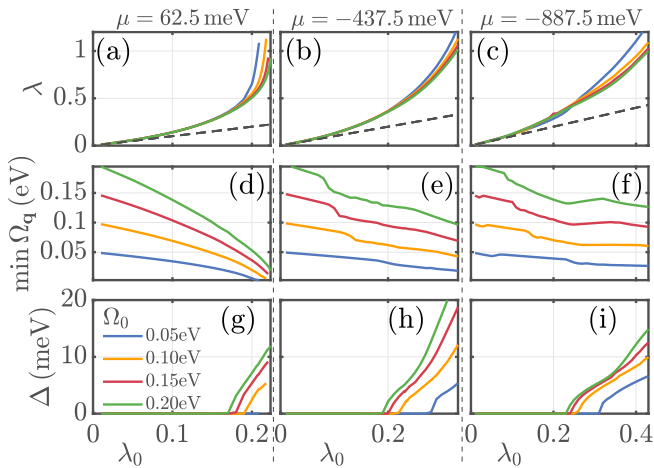}
	\caption{Results for the renormalized coupling strength $\lambda$ (upper row), minimum renormalized phonon frequency (middle row) and maximum superconducting gap (lower row), calculated for our 2D systems. Results shown in the first, second and third column are computed with the orange, green and purple dispersion of Fig.\,\ref{energies}(a), respectively. Different colors correspond to bare phonon frequencies $\Omega_0$ as written in the legend of panel (g).}\label{trends2D}
\end{figure}

As a first step we want to understand how the input parameters influence our self-consistent results. For this purpose we solve the Eliashberg equations in 2D as function of $\lambda_0$ at $T=20\,\mathrm{K}$ for various bare frequencies $\Omega_0$. We show our results for dispersions (1), (2) and (3) of Fig.\,\ref{energies}(a), respectively, in the first, second, and third column of Fig.\,\ref{trends2D}. In each of these columns we show results for different initial frequencies $\Omega_0$ with color code as indicated in the legend of panel (g). The upper row shows the renormalized coupling strength as function of $\lambda_0$, where we add as guide for the eye the case $\lambda=\lambda_0$ as gray dashed line. The middle row shows the minimum renormalized phonon frequency, again as function of bare input coupling. The maximum superconducting gap, defined by $\Delta=\underset{\mathbf{k}\in \mathrm{BZ}}{\mathrm{max}}\,\Delta_{\mathbf{k},m=0}$, is shown in the lower row.

In Fig.\,\ref{trends2D}(a-c) we observe the well-known behavior of the renormalized coupling with increasing $\lambda_0$ \cite{Scalettar1989,Bauer2011,Esterlis2018}, i.e., $\lambda$ goes towards a divergence, which we define to occur at $\lambda_0^{\star}$. As is apparent in panels (d-f), when $\lambda_0\rightarrow\lambda_0^{\star}$ the minimal phonon frequency vanishes, indicating a lattice instability. The average frequency on the other hand decreases approximately linearly with $\lambda_0$ (not shown). From Fig.\,\ref{energies}(a) we know that the FS nesting condition is met less accurately as we go from the orange (1) to the purple dispersion (3). The phonon self-energy exhibits less coherent contributions for the shallow energy band (3), which in turn renormalizes the phonon propagator to a smaller extend. For this reason we find increasing values of $\lambda_0^{\star}$ as we go from the first to the third column in Fig.\,\ref{trends2D}. In line with this observation we find the fastest decrease of minimum phonon frequencies as function of $\lambda_0$ in panel (d), and the slowest in panel (f).

From each of Fig.\,\ref{trends2D}(g-i) we observe that the superconducting gap opening with respect to $\lambda_0$ depends on $\Omega_0$. An increase in the initial phonon frequency reduces the minimal bare coupling strength needed for a finite superconducting gap. Further, comparing these three panels shows that the onset of superconductivity depends also on the electronic dispersion. This observation can again be understood in terms of changed nesting conditions, see the discussion above. When considering the superconducting gap as function of $\lambda$ (instead of $\lambda_0$), the difference in the onset of $\Delta\neq0$ among results calculated for our three 2D electron dispersions becomes smaller. The onset of superconductivity with respect to $\lambda_0$ is discussed in more detail later in this Section, and further analysis of the superconducting state in connection with phonon renormalization is provided in Section \ref{scTc}.

Including the back reaction of electrons on the phonon spectrum, via the Feynman diagram shown in Fig.\,\ref{feynman}(b), leads to a decrease in the magnitude of frequencies, which is a well known behavior commonly referred to as phonon softening\,\cite{Allen1983}. This phenomenon has been discussed especially in context of the 2D Holstein model\,\cite{Scalettar1989,Marsiglio1990,Meyer2002,Dee2019}, and it marks tendencies of the system to develop a Charge Density Wave instability. The question of whether phonon softening is favorable for superconductivity or suppresses $T_c$ is discussed in further detail in Section \ref{scTc}.

\begin{figure}[h!]
	\centering
	\includegraphics[width=1\linewidth]{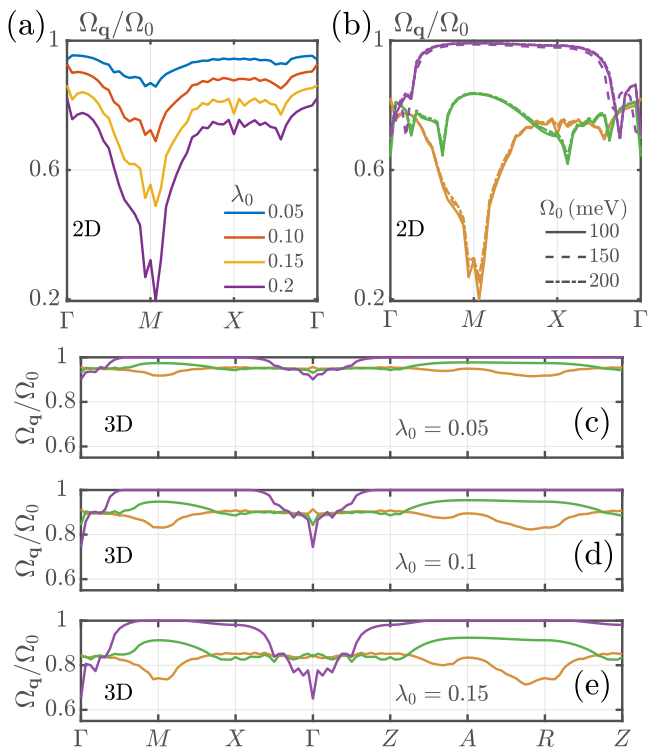}
	\caption{Momentum dependent phonon frequencies, normalized to the initial choice of $\Omega_0$. Calculations were performed at $T=20\,\mathrm{K}$. (a) Results for different bare couplings $\lambda_0$ as written in the legend, computed for $\Omega_0=100\,\mathrm{meV}$ and the orange dispersion {(1)} of Fig.\,\ref{energies}(a). (b) Different colors correspond to the choice of 2D electron dispersions, compare Fig.\,\ref{energies}(a). With $\lambda_0=0.2$, the three different line styles represent varying choices of bare frequency $\Omega_0$ as written in the legend. (c), (d), (e) Renormalized phonon frequencies along high-symmetry lines of the BZ. Different color{s}  correspond to 3D electron energies as shown in Fig.\,\ref{energies}(c). Bare couplings are chosen as $\lambda_0=0.05$, $\lambda_0=0.1$ and $\lambda_0=0.15$.}\label{freq}
\end{figure}

To first show the general effect, let us consider the orange dispersion (1) in Fig.\,\ref{energies}(a) and a bare frequency $\Omega_0=100\,\mathrm{meV}$. After self-consistently solving the Eliashberg equations in 2D at $T=20\,\mathrm{K}$, we calculate the renormalized phonon spectrum via Eq.\,(\ref{omegatilde}). In Fig.\,\ref{freq}(a) we plot our result for $\Omega_{\mathbf{q}}/\Omega_0$ along high-symmetry lines of the BZ for various coupling strengths as indicated in the legend. In the limit of small $\lambda_0$, see blue curve, the renormalization effects are relatively minor, i.e.\ $\Omega_{\mathbf{q}}/\Omega_0$ stays close to unity throughout the BZ. The biggest phonon softening occurs at $\mathbf{q}\simeq M$ because the FS is relatively well nested at this exchange momentum. As we increase $\lambda_0$ we confirm the leading instability to be at $M$, since the smallest ratio of renormalized to bare frequencies is observed at this $\mathbf{q}$, see Fig.\,\ref{freq}(a). Additionally, phonon softening occurs throughout the whole BZ, so we observe $\Omega_{\mathbf{q}}<\Omega_0\,\forall\mathbf{q}$ for any finite $\lambda_0$. 

We can understand the observed decreases in $\Omega_{\mathbf{q}}$ by expressing the phonon self-energy as $\Pi_{\mathbf{q},l} = -g_{\mathbf{q}}^2\chi^0_{\mathbf{q},l}$, where $\chi^0_{\mathbf{q},l}$ is the charge susceptibility\,\cite{Dee2019}. Inserting into Eq.\,(\ref{omegatilde}) and using $g_{\mathbf{q}}=g_0$ gives $\Omega_{\mathbf{q}}/\Omega_0 = \sqrt{\Omega_0^2 - 2\Omega_0g_{0}^2\chi^0_{\mathbf{q},l=0}}/\Omega_0 = \sqrt{1 - 2g_{0}^2\chi^0_{\mathbf{q},l=0}/\Omega_0}$. Therefore the anisotropy in the renormalized frequencies comes solely from the susceptibility, which in turn is dominated by contributions due to FS nesting. In addition we note that $g_0^2/\Omega_0\propto\lambda_0$, therefore the phonon softening is expected to increase with coupling strength, a trend confirmed by our calculations shown in Fig.\,\ref{freq}(a).

To further examine the influence of nesting on the phonon renormalization we test the three dispersions as shown in Fig.\,\ref{energies}(a). For each $\xi_{\mathbf{k}}$ we choose initial frequencies $\Omega_0 =100,~150$, or $200\,\mathrm{meV}$ and a coupling strength of $\lambda_0=0.2$. Our results for $\Omega_{\mathbf{q}}/\Omega_0$ are plotted in Fig.\,\ref{freq}(b), again as function of exchange momentum $\mathbf{q}$ along high symmetry lines. The color code is identical to Fig.\,\ref{energies}(a), and we use varying line styles, as written in the legend, to show results computed for different $\Omega_0$. 

As first observation we find that our results are to a good approximation independent of the choice of bare phonon frequency. This is reflected in the fact that all three curves (different line styles) for any of the electron dispersions fall almost precisely on top of each other. Therefore we conclude that $\Omega_{\mathbf{q}}/\Omega_0$ is a direct function of $\lambda_0$ (see panel (a)) but not of $\Omega_0$. This can be seen by expressing the above functional form as $\Omega_{\mathbf{q}}/\Omega_0 = \sqrt{1 - \lambda_0\chi^0_{\mathbf{q},l=0}/N_0}$ via Eq.\,(\ref{lambda0}). Notably, there is still an implicit dependence on $\Omega_0$ hidden in
\begin{align}
\chi^0_{\mathbf{q},l}= - T \sum_{\mathbf{k},m} \mathrm{Tr}\big[ \hat{G}_{\mathbf{k},m} \hat{\rho}_3 \hat{G}_{\mathbf{k}+\mathbf{q},m+l} \hat{\rho}_3 \big]
\end{align} 
due to the self-consistency of our approach, which is why the curves are not precisely equivalent.

When comparing results computed for different electron dispersions we find much bigger effects in the renormalized phonon spectrum. Starting with the orange curves of Fig.\,\ref{freq}(b), we find the most pronounced phonon softening at $\mathbf{q}=M$. This case is already discussed above, and can be explained by well satisfied nesting conditions at this wavevector. The green curves show larger values for $\Omega_{\mathbf{q}}/\Omega_0$, i.e. the renormalization effects are less pronounced compared to results shown in orange. From the FS properties, see Fig.\,\ref{energies}(b), we know that the exchange momentum is no longer close to $M$ when focusing on the green curve, rather $\mathbf{q}$ lies in between $\Gamma$ and $M$ or $X$. This nesting property is directly translated into results of Fig.\,\ref{freq}(b), where the softest phonons are found along $\Gamma-M$ and $X-\Gamma$. Turning now to the most shallow energy band, the purple lines (3) in Figs.\,\ref{freq}(b) and \ref{energies}(a), the phonon frequencies $\Omega_{\mathbf{q}}$ are almost as large as their respective $\Omega_0$, for $\mathbf{q}$ between $M$ and $X$. Since the FS is a very small circle at the center of the BZ, it is not surprising that the susceptibility peaks only at small wavevectors. Therefore we observe softer phonons around $\Gamma$ in these results.

In the discussion above we did identify FS nesting as an important factor when considering phonon softening and renormalized couplings in 2D. We now want to explore this aspect also in 3D systems, using electron energies from Fig.\,\ref{energies}(c). Fixing $\Omega=100\,\mathrm{meV}$, we solve the Eliashberg equations for three different couplings strengths, $\lambda_0=0.05$, $\lambda_0=0.1$ and $\lambda_0=0.15$, at $T=20\,\mathrm{K}$. The corresponding results for the phonon spectrum are shown in Figs.\,\ref{freq}(c), \ref{freq}(d), and \ref{freq}(e), respectively. Each of these panels contains results for all three electron dispersions tested, where we adopt the color code of Fig.\,\ref{energies}(c-f). The first observation is similar to the 2D case, i.e. the phonons become softer as we increase the coupling strength. This can be seen by comparing similarly colored curves among different panels of Fig.\,\ref{freq}(c-e). Further it is apparent that the phonon softening is generally less pronounced and coherent
in 3D. The orange curves (1) in each panel show a relatively small tendency for phonon softening, while the smallest $\Omega_{\mathbf{q}}/\Omega_0$ are detected at $M$ and $R$. Effects are even less prominent for the green lines (2) of each panel, where, for couplings up to $\lambda_0=0.15$, $\Omega_{\mathbf{q}}/\Omega_0$ stays relatively close to unity throughout the BZ. For the most shallow energy dispersion (3), which is represented by purple lines in Fig.\,\ref{freq}(c-e), we find the results with highest anisotropy. Throughout most parts of the BZ the phonon spectrum is to a good approximation not renormalized, but $\Omega_{\mathbf{q}}/\Omega_0$ decreases strongly around $\Gamma$. This behavior is qualitatively similar to the 2D case shown in Fig.\,\ref{freq}(b), in contrast to the other curves shown here (green, orange).

From the above discussion we learn that reduced nesting properties lead to less phonon softening in 3D, compared to the 2D situation. However, very shallow electron dispersions are an exception to this trend because we observe comparable results of $\Omega_{\mathbf{q}}/\Omega_0$ in 3D and 2D. The reason lies in the special role of the $\Gamma$ point: When considering a spherical FS, $\mathbf{q}\sim\Gamma$ is the only point in the BZ with which FS parts can be connected without dependence on the angle of wave vector $\mathbf{q}$, hence the susceptibility response in 3D can develop similarly coherent contributions at $\Gamma$ as in 2D. 

We now turn to a more detailed discussion of the renormalized coupling strength $\lambda$, which, as we argue below, does similarly as the phonon softening strongly depend on FS nesting properties, and hence follows different trends in 2D and 3D. As stated before, the coupling diverges as we increase $\lambda_0\rightarrow\lambda_0^{\star}$, with $\lambda_0^{\star}$ marking a lattice instability. Recalling the definition $\lambda = \langle \langle \lambda_{\mathbf{k}-\mathbf{k}'}\rangle_{\mathbf{k}\in FS}\rangle_{\mathbf{k}'\in FS}$, we combine Eqs.\,(\ref{lambdatilde}) and (\ref{invD}) to write the momentum dependent coupling as
\begin{align}
\lambda_{\mathbf{q}} = N_0 g_0^2 \left[ \frac{\Omega_0}{2} + \Pi_{\mathbf{q},l=0} \right]^{-1} ~, \label{lambdaqestimate}
\end{align}
where we used $g_{\mathbf{q}}=g_0$. With $\Pi_{\mathbf{q},l=0}=-g_0^2 \chi_{\mathbf{q},l=0}$ and $\lambda_0 = 2g_0^2N_0/\Omega_0$, we get 
\begin{align}
\lambda &= \langle \langle \lambda_{\mathbf{k}-\mathbf{k}'}\rangle_{\mathbf{k}\in FS}\rangle_{\mathbf{k}'\in FS} \nonumber \\
&= \Big\langle \Big\langle \frac{N_0 \lambda_0}{N_0 - \lambda_0 \chi_{\mathbf{q},l=0}} \Big\rangle_{\mathbf{k}\in FS}\Big\rangle_{\mathbf{k}'\in FS} . \label{lambdaestimate}
\end{align}
We model Eq.\,(\ref{lambdaestimate}) by introducing the fitting function
\begin{align}
\lambda \sim a_{\lambda} \cdot \frac{N_0 \lambda_0}{N_0 - b_{\lambda} \cdot \lambda_0} , \label{lambdafit}
\end{align}
where $a_{\lambda}$ reflects the magnitude of the coupling, and $b_{\lambda}$ is a measure of the influence due to $\chi_{\mathbf{q},l=0}$. Needless to say, Eq.\,(\ref{lambdafit}) is an approximation, since we neglect the momentum dependence of the susceptibility. As mentioned earlier, the renormalized coupling diverges at a critical choice $\lambda_0^{\star}$ of the bare coupling strength. We can find an estimate of this quantity from the denominator in Eq.\,(\ref{lambdafit}), namely, when $(N_0 - b_{\lambda} \cdot \lambda_0)\rightarrow0$ we can write 
\begin{align}
\lambda_0^{\star}\simeq \frac{N_0}{b_{\lambda}} ~. \label{lcrit}
\end{align}

\begin{figure}[b!]
	\centering
	\includegraphics[width=1\linewidth]{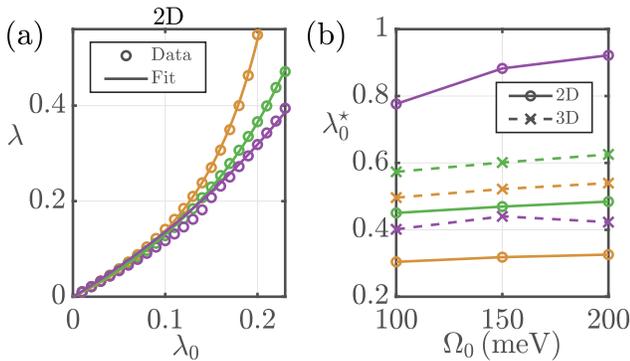}
	\caption{ (a) Renormalized coupling strength as function of $\lambda_0$, calculated in 2D for $\Omega_0=100\,\mathrm{meV}$ and $T=20\,\mathrm{K}$. Open circles represent our self-consistent results, solid lines are obtained from Eq.\,(\ref{lambdafit}). (b) Critical couplings $\lambda_0^{\star}$ as function of $\Omega_0$, shown for different dispersions in 2D and 3D.
	{The color code used here in (a), (b) reflects the choice of electron dispersion according to Figs.\,\ref{energies}(a) and \ref{energies}(c). Results obtained for 2D (3D) are shown via solid (dashed) lines.}}\label{scaleLambda}
\end{figure}

In Fig.\,\ref{scaleLambda}(a) we show our self-consistent results for $\lambda$ as function of $\lambda_0$ as open circles, for $\Omega_0=100\,\mathrm{meV}$, $T=20\,\mathrm{K}$ and the three 2D electron dispersions of Fig.\,\ref{energies}(a) (similar color code). The fitted behavior as described in Eq.\,(\ref{lambdafit}) is shown as solid lines for each data set. As is apparent from this graph, the renormalized coupling strength can be modeled with the above functional form to a very good approximation. Repeating the same procedure for $\Omega_0=150\,\mathrm{meV}$ and $\Omega_0=200\,\mathrm{meV}$, and for all three 3D electron energies leads to the curves shown in Fig.\,\ref{scaleLambda}(b), where we plot the critical bare coupling as function of input frequency. Solid (dashed) lines and open circles (crosses) represent trends in 2D (3D), while colors are again corresponding to {the electron dispersions} Figs.\,\ref{energies}(a) and \ref{energies}(c). 

From all curves of Fig.\,\ref{scaleLambda}(b) we learn that $\lambda_0^{\star}$ is to first order approximation independent of the initial phonon frequency, which goes in line with observations from Fig.\,\ref{freq}(b). Further, our 2D results reflect the expected behavior with respect to nesting conditions: The orange dispersion (1) of Fig.\,\ref{energies}(a) is relatively well nested, therefore the susceptibility shows large contributions that are reflected in the fitting constant $b_{\lambda}$. Consequently, the critical input coupling, compare Eq.\,(\ref{lcrit}), is a small number. As we proceed  to the green (2), and eventually the purple (3) electron dispersion, the DOS decreases, while at the same time the nesting becomes less coherent. Evidently, the latter effect is more important since the susceptibility values (i.e.\ $b_{\lambda}$) decrease faster than $N_0$, so that the resulting $\lambda_0^{\star}$ grows larger. Turning to the dashed curves of Fig.\,\ref{scaleLambda}(b) we find a different behavior. The values for $\lambda_0^{\star}$ are far less susceptible for changes in the electron dispersion. Further, we find smaller values for $b_{\lambda}$ in 3D than in 2D, which 
clarifies that nesting in 3D is less important. An exception to this is the result drawn in purple, representing the most shallow electron dispersion {(3)}. As discussed in connection to Fig.\,\ref{freq}(c-e), this stems from the special role of nesting at $\mathbf{q}\sim\Gamma$, which is developed to a larger extend in 3D than in 2D. For this reason (large value of $b_{\lambda}$) the purple dashed line falls not only below the two other curves for 3D systems, but lies also lower than the purple solid line obtained for the most shallow 2D electron dispersion (3). 

Note, that the results of Fig.\,\ref{scaleLambda}(b) are obtained by fitting $\lambda$ in the non-critical range of the input coupling, i.e.\ for $\lambda_0$ significantly smaller than $\lambda_0^{\star}$. Therefore the reported values of $\lambda_0^{\star}$ are to be interpreted as qualitative trends, and should not be taken as precise numbers. Performing a calculation at $\lambda_0\sim\lambda_0^{\star}$ is numerically very difficult, because the input $\xi_{\mathbf{k}}-\delta\mu$ has to be gradually adjusted so as to keep the electron filling at the desired level. If, however, the interplay of $\delta\mu$, $\Omega_0$ and $\lambda_0$ is such that the input coupling is above the respective $\lambda_0^{\star}$, the self-consistent Eliashberg loop never converges.

\begin{figure}[b!]
	\centering
	\includegraphics[width=1\linewidth]{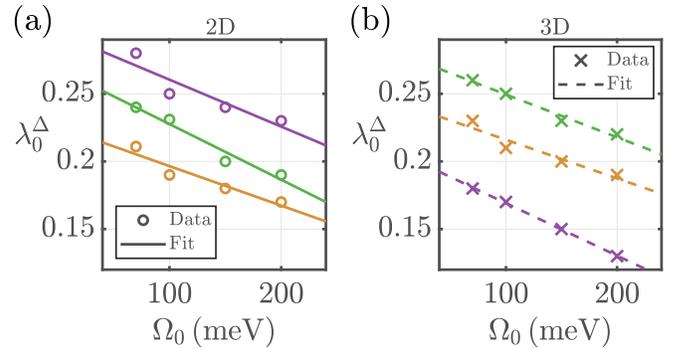}
	\caption{Bare coupling strength $\lambda_0^{\Delta}$, corresponding to the onset of superconductivity, as function of input frequency {$\Omega_0$}. Results are computed for $T=20\,\mathrm{K}$, and the color code corresponds to Fig.\,\ref{energies}. (a) Results for 2D electron dispersions. (b) Results for 3D electron dispersions.}\label{scaleLambdaDelta}
\end{figure}

We end this section by looking into the bare coupling strength $\lambda_0^{\Delta}$, at which the onset of superconductivity occurs, i.e., the smallest $\lambda_0$ at which $\Delta_{\mathbf{k},m}\neq0$. As is apparent from Fig.\,\ref{trends2D}, $\lambda_0^{\Delta}$ depends both on properties of the electron dispersion and the phonon frequency, hence we want to examine this quantity closer in 2D and 3D. We choose phonon frequencies $\Omega_0$ as 70, 100, 150, or 200 $\mathrm{meV}$, $T=20\,\mathrm{K}$ and use electron energies from Fig.\,\ref{energies}(a) and Fig.\,\ref{energies}(c) to show computed results for $\lambda_0^{\Delta}$ as function of $\Omega_0$ in Fig.\,\ref{scaleLambdaDelta}(a) and Fig.\,\ref{scaleLambdaDelta}(b), for 2D and 3D, respectively. Open circles (crosses) represent our data, while solid (dashed) lines are obtained by a linear fit of the 2D (3D) data. The color code corresponds to choices of electron dispersion according to Fig.\,\ref{energies}. {For each of the curves, both in 2D and 3D, we observe a linear decrease in $\lambda_0^{\Delta}$ with increasing frequency. This stems from the fact that renormalized frequencies $\Omega_{\mathbf{q}}$ are growing with $\Omega_0$, enhancing the tendencies of the system to form a superconducting state.} 

When comparing results for $\lambda_0^{\Delta}$ with respect to shallowness and nesting, we find similar trends with those already observed in Figs.\,\ref{freq} and \ref{scaleLambda}. In the 2D case, Fig.\,\ref{scaleLambdaDelta}(a), the orange dispersion (1) has the best nesting conditions, resulting in a large susceptibility at the nesting wave vector. This in turn leads to a relatively large renormalized coupling strength $\lambda$. By increasing shallowness of the electron energies, hence considering the green (2) and purple (3) lines, we obtain decreasing magnitudes of $\chi_{\mathbf{q},l}$, which results in smaller coupling strengths. Hence, to achieve similar values of $\lambda$ as in the well nested case, $\lambda_0$ has to be increased. For 3D systems nesting is generally less coherent, therefore the orange (1) and green (2) curves in Fig.\,\ref{scaleLambdaDelta}(b) show larger values for $\lambda_0^{\Delta}$ than in Fig.\,\ref{scaleLambdaDelta}(a), with similar reasoning. The very shallow $\xi_{\mathbf{k}}$ (3), represented by the purple line, is again the exception due to exhibiting coherent nesting behavior at $\Gamma$, which is why only small couplings are needed to induce superconductivity.

\section{Critical temperatures}\label{scTc}

\begin{figure}[b!]
	\centering
	\includegraphics[width=1\linewidth]{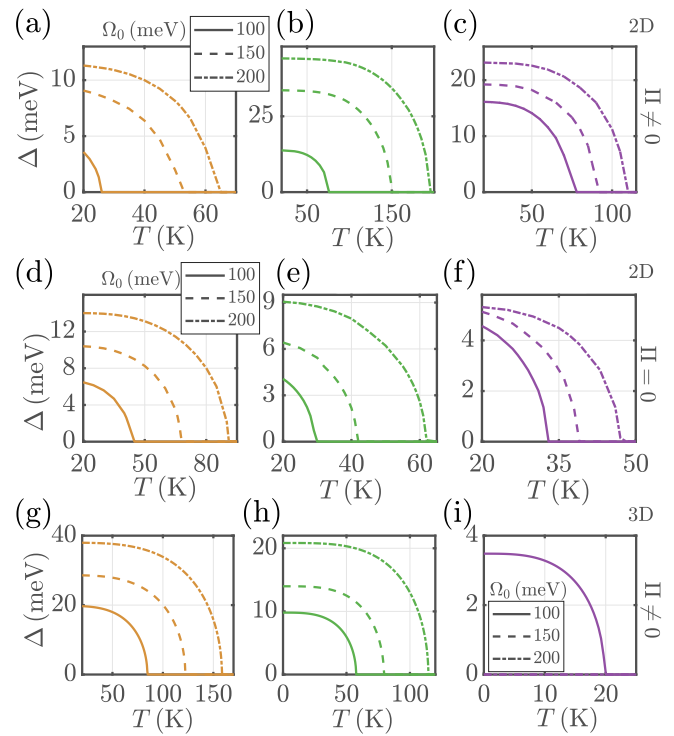}
	\caption{Maximum superconducting gap as function of temperature with color code corresponding to the electron dispersions of Fig.\,\ref{energies}(a) and Fig.\,\ref{energies}(c). The different line styles represent the choices of initial frequency $\Omega_0$ as written in the legends. For the size of input couplings $\lambda_0$ see main text. (a-c) Results for $\Delta(T)$ for the 2D electron systems. (d-f) Self-consistent superconducting gaps for our 2D dispersions, obtained under the assumption $\Pi_{\mathbf{q},l=0}=0$, using $\lambda_0=\lambda=0.45$. (g-i) Results, including the phonon renormalization, for the 3D systems of Fig.\,\ref{energies}(c), with coupling strengths as described in the main text.}\label{gap}
\end{figure}

Let us now turn to the superconducting critical temperature $T_c$ that can be determined by following the maximum superconducting gap with $T$. We define $T_c$ as the smallest temperature at which $\Delta$ vanishes and use the same electron dispersions as before, see Figs.\,\ref{energies}(a) and \ref{energies}(c). Starting in 2D, we test the initial frequencies $\Omega_0$ being 100, 150, or 200 $\mathrm{meV}$, which are represented in Fig.\,\ref{gap}(a-c) by solid, dashed, and dotted-dashed lines, respectively. All panels show the maximum superconducting gap as function of temperature, with color code corresponding to the choice of electron dispersion from Fig.\,\ref{energies}(a). For each $\xi_{\mathbf{k}}$ and $\Omega_0$ we take a relatively large value for $\lambda_0$ close to the lattice instability, so as to maximize the renormalized coupling and critical temperature.

For panel (a) we take the parameter choices $(\Omega_0,\lambda_0)=(100\,\mathrm{meV},0.205),$ $(150\,\mathrm{meV},0.215)$, and $(200\,\mathrm{meV},0.22)$. The resulting critical temperatures lie between $26\,\mathrm{K}$ and $65\,\mathrm{K}$. Next we turn to panel (b): The Fermi surface is smaller than in the case before, but the electron band (2) is not yet very shallow. We use $(\Omega_0,\lambda_0)=(100\,\mathrm{meV},0.34)$, $(150\,\mathrm{meV},0.41)$ and $(200\,\mathrm{meV},0.42)$ for the solid, dashed and dotted-dashed curves. As easily observed, $T_c$ drastically grows when compared to Fig.\,\ref{gap}(a), with the largest critical temperature almost reaching $200\,\mathrm{K}$. When we finally go to the results for a shallow band (3), panel (c), $T_c$ ranges between $\sim80\,\mathrm{K}$ and $\sim110\,\mathrm{K}$. Here the curves are produced by choosing  $(\Omega_0,\lambda_0)=(100\,\mathrm{meV},0.54)$, $(150\,\mathrm{meV},0.53)$, and $(200\,\mathrm{meV},0.55)$.

{The here observed results can be explained by an intuitive picture: The dispersion corresponding to panel (a) exhibits the best nesting conditions, which leads to enhanced renormalization of the phonon propagator (see also Section \ref{scSoftCoup}). This means that a substantial part of the available coupling in the system is `used' for renormalizing the phonon frequency, leaving less coupling available for Cooper pairing and therefore resulting in a reduction of $\Delta$ and $T_c$. In the opposite limit of {the} shallow band (3), Fig.\,\ref{gap}(c), most of the initial coupling strength is available to form the superconducting state, because the nesting, and therefore the renormalization of the phonon spectrum is comparatively minor. Consequently $T_c$ is enhanced when comparing to Fig.\,\ref{gap}(a). However, due to the shallowness of the electron band (3) we face a reduced FS area and DOS, therefore the $T_c$'s are larger than in panel (a), but not maximized.} 

The results shown in panel (b) can be seen as example of how to achieve the highest possible critical temperatures. The electron dispersion (2) exhibits relatively bad nesting conditions, while still not being in the limit of a very shallow band. For achieving a large magnitude of $\Delta$ (and therefore $T_c$) a substantial renormalized coupling $\lambda$ is required, combined with a large FS density of states. The system of panel (b) lies in an ideal intermediate regime, where the balance between phonon renormalization and electron band shallowness is kept. The FS nesting condition is good enough for achieving a large $\lambda$, but not as ideal as in Fig.\,\ref{gap}(a) so as to drive the system towards a lattice instability before superconductivity can build up. Additionally the FS DOS is sufficiently large, in contrast to panel (c), so as to boost values for $T_c$.

For explicitly testing this interpretation we perform additional calculations for the same frequencies $\Omega_0$ and 2D electronic dispersions as before, but under the approximation $\Pi_{\mathbf{q},l}=0$, i.e., considering only the bare phonon propagator $D_{\mathbf{q},l}=D^0_{\mathbf{q},l}$. The coupling strength is fixed at $\lambda_0=0.45$. Our results for the maximum superconducting gap as function of temperature are shown in Fig.\,\ref{gap}(d-f), with similar color codes and line styles as in Fig.\,\ref{gap}(a-c). For each panel of Fig.\,\ref{gap}(d-f) we find that an increase in $\Omega_0$ leads to larger gap magnitudes and critical temperatures, as is expected. When comparing the results for different electron dispersions, we observe that $\Delta$ and $T_c$ decrease as we go from high (panel (a)) to low (panel (c)) FS DOS.

From these results for $\Pi_{\mathbf{q},l}=0$ we recover the standard BCS result  that the superconducting critical temperature depends on the phonon frequency and FS DOS.
This proves our point that we made in connection to Fig.\,\ref{gap}(a-c), because a finite $\Pi_{\mathbf{q},l}$ will decrease all values of Fig.\,\ref{gap}(d-f) to an extend that depends only on the nesting condition. Therefore there exists an ideal balance between phonon renormalization and FS DOS, which maximizes the critical temperature. We note that a comparison between Figs.\,\ref{gap}(a-c) and \ref{gap}(d-f) is done here only in a qualitative way, since a rigorous matching of system parameters is not straight forward.

Let us now turn to 3D systems, taking electron energies $\xi_{\mathbf{k}}$ from Fig.\,\ref{energies}(c). The maximum superconducting gap as function of temperature is shown in Fig.\,\ref{gap}(g-i), where we test again three phonon frequencies for each dispersion (see legend). Each curve is obtained by imposing a respective input coupling $\lambda_0$ that is close to the lattice instability. In panel (g) we plot our self-consistent solutions for $(\Omega_0,\lambda_0)=(100\,\mathrm{meV},0.383)$, $(150\,\mathrm{meV},0.4)$ and $(200\,\mathrm{meV},0.42)$, using the orange electron dispersion (1) of Fig.\,\ref{energies}(c). As before we find a clear enhancement in $\Delta$ and $T_c$ with increased phonon frequency. Results for both maximum superconducting gap and critical temperature become comparatively smaller in Fig.\,\ref{gap}(h), $(\Omega_0,\lambda_0)=(100\,\mathrm{meV},0.31)$, $(150\,\mathrm{meV},0.31)$, $(200\,\mathrm{meV},0.33)$, where the green dispersion of Fig.\,\ref{energies}(c) is used. The reason for this trend is a decrease in FS DOS, combined with a less important role of nesting properties in 3D systems. Finally, we show $\Delta(T)$ for the purple 3D dispersion (3), see Fig.\,\ref{energies}(c), in Fig.\,\ref{gap}(i). Here we found no superconductivity for $(\Omega_0,\lambda_0)=(200\,\mathrm{meV},0.15)$ and $(150\,\mathrm{meV},0.12)$ down to $10\,\mathrm{K}$. The gap size and $T_c$ for $(\Omega_0,\lambda_0)=(100\,\mathrm{meV},0.1)$ are substantially smaller than in panels (g) and (h). These results can be explained under the light of an extremely small FS DOS of the purple electron dispersion {(3)}. Additionally, large parts of the available coupling are `used' for renormalizing the phonon spectrum, due to well enhanced nesting conditions at $\Gamma$, see also discussions in Section \ref{scSoftCoup}.  

\begin{figure}[h!]
	\centering
	\includegraphics[width=1\linewidth]{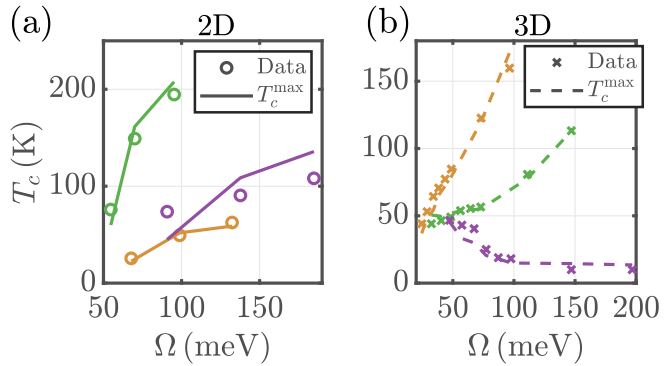}
	\caption{ Influence of phonon renormalization on $T_c$ in 2D and 3D systems. Open circles and crosses represent computed values for $T_c$ as function of renormalized frequency $\Omega$, where the color code of Fig.\,\ref{energies}(a) (2D) and Fig.\,\ref{energies}(c) (3D) is used for panels (a) and (b). The couplings have been chosen close to the lattice instability. In the main text we describe how to obtain the approximate upper bounds for $T_c$ as shown by the dashed lines.}\label{tcbound}
\end{figure}

Above we have already encountered some aspects concerning the superconducting $T_c$, in both 2D and 3D systems. In the following we propose a scaling of the critical temperature, which can also serve as approximate upper bound. In Fig.\,\ref{tcbound} we show our maximally possible values for $T_c$ as function of renormalized phonon frequency, adopting the color code from Fig.\,\ref{energies}. Our computed results for 2D and 3D systems are respectively shown in panel (a) (open circles) and panel (b) (crosses). For finding these critical temperatures we choose initial frequencies $\Omega_0\in[50,200]\,\mathrm{meV}$, and pick the largest $\lambda_0<\lambda_0^{\star}$ for which a solution can be stabilized.

Our results show that the maximum $T_c$ obtained here always increases as function of renormalized frequency $\Omega=\langle\Omega_{\mathbf{q}}\rangle_{\mathbf{q}\in\mathrm{BZ}}$ for 2D systems. However, in three spatial dimensions this trend is not observed for our most shallow dispersion (3), where $T_c<10\,\mathrm{K}$ for $\Omega\gtrsim150\,\mathrm{meV}$. For modeling the functional behavior found in Fig.\,\ref{tcbound}, it is worthwhile considering trends reported by other authors. In Ref.\,\cite{Esterlis2018_2}, Esterlis \textit{et al.} proposed that $T_c^{\mathrm{max}}\propto \Omega$ constitutes a reasonable upper bound for the critical temperature. However, our results do not appear linear in the renormalized phonon frequency, which is why this scaling might be a too crude approximation. Another proposal was made by Moussa and Cohen, stating that $T_c^{\mathrm{max}}\propto\sqrt{\Omega_0^2-\Omega^2}$\,\cite{Moussa2006}. Imposing this form does indeed lead to satisfying agreement with our data in 2D, but fails to capture the decrease in critical temperature observed for the purple line in Fig.\,\ref{tcbound}(b). We find that the best fit to our numerical data is given by the scaling expression
\begin{align} 
k_B T_c^{\mathrm{max}}\propto\lambda\sqrt{\Omega_0^2-\Omega^2}  ~, \label{fittc}
\end{align}
which is shown as dashed lines in Fig.\,\ref{tcbound}. As is apparent from comparing our data with outcomes of Eq.\,(\ref{fittc}), the functional dependence in Fig.\,\ref{tcbound} is quite  accurately captured. Especially for 3D systems the inclusion of $\lambda$ is crucial to mimic the observed trends, while the functional dependence proposed in Ref.\,\cite{Moussa2006} would suffice for the 2D cases. 

It is important to notice the difference between Eq.\,(\ref{fittc}) and the more simplified scaling law of Ref.\,\cite{Moussa2006}, since $\lambda$ is not constant with respect to $\Omega$. To show this explicitly, we can solve Eq.\,(\ref{omegatilde}) for the zero-frequency phonon self-energy, yielding $\Pi_{\mathbf{q},l=0}=(\Omega_{\mathbf{q}}^2-\Omega_0^2)/(2\Omega_0)$. We insert this expression into Eq.\,(\ref{lambdaqestimate}) and, as an approximation replace $\Omega_{\mathbf{q}}$ by $\Omega$, so that
\begin{align}
k_{B} T_c^{\rm max} &\sim 2N_0 g_0^2 \frac{\Omega_0}{\Omega^2}  \sqrt{\Omega_0^2-\Omega^2}   ~. \label{fittcomega}
\end{align}

The interpretation of Eqs.\,(\ref{fittc}) and (\ref{fittcomega}) is rather intuitive: The highest possible value of $T_c$ depends directly on the renormalized coupling strength in the system. Additionally, well developed phonon softening is advantageous for maximizing the critical temperature, i.e., having $\Omega\ll\Omega_0$. These two effects are not decoupled from each other, since both depend on nesting properties and the DOS. For 2D we already observed the trends in Fig.\,\ref{gap}(a-c); the green curve in Fig.\,\ref{tcbound}(a) represents the ideal compromise between magnitude of the DOS (purple curve being suppressed because DOS is too low) and nesting properties (orange curve being suppressed because phonon softening is too strong). Going from the orange to the green curve in panel (b) of Fig.\,\ref{tcbound} reflects the decrease of $N_0$, compare Fig.\,\ref{energies}. The most shallow 3D system exhibits additionally coherent nesting at $\Gamma$, so as to produce negative phonon frequencies before superconductivity can fully develop. Summarizing, we obtain numerically that the effect of phonon softening is in general favorable for superconductivity, provided that the un-renormalized phonon frequency $\Omega_0$ is sufficiently large, such that the system is not developing a lattice instability before superconductivity can fully build up.

\section{Conclusion}\label{scConclusion}

In this work we have investigated the details of Eliashberg theory including self-consistent phonon renormalization on a model basis. We worked out the similarities and differences between 2D and 3D systems, identifying nesting properties and FS DOS as key aspects. From those follow directly the trends for phonon softening, electron-phonon renormalization and the critical temperature. Our study therefore constitutes an extensive overview of possible effects of phonon renormalization in Eliashberg theory under Migdal's approximation.

Our calculations show that the maximally possible $T_c$ in 2D results from a delicate interplay between FS nesting and DOS. If the size of the FS (and hence the DOS) is too small, no strongly increased values for $T_c$ are found. On the other hand, very coherent nesting, which in our model is associated with a big FS, drives the system too quickly towards a lattice instability with increasing coupling strength. Maximal values for the critical temperature are therefore found in an intermediate regime. In three spatial dimensions, nesting, and therefore the renormalization of the phonon spectrum becomes less coherent. Consistently we find that the magnitude of $T_c$ is mainly dictated by the size of the FS.

In Section \ref{scSoftCoup} we have shown that there exists an inherent bound $\lambda_0^{\star}$ on the maximal coupling strength, which is not to be confused with critical values of $\lambda_0$ in other works. Chubukov {\em et al.}\ use the notation $\lambda_{\mathrm{cr}}$ to describe the border of applicability of Eliashberg theory with respect to the electron-phonon coupling strength\,\cite{Chubukov2020}. This border is found by comparing outcomes from Eliashberg theory with QMC simulations in the normal state, so that for $\lambda>\lambda_{\mathrm{cr}}$ the two theories do no longer produce similar results. The same border of applicability is denoted $\lambda^{\star}$ in Ref.\,\cite{Esterlis2018}, which, again is not the same coupling strength as our $\lambda_0^{\star}$ corresponding to a lattice instability. 

As mentioned before, various authors agree on the existence of a maximal coupling strength up to which Eliashberg theory produces accurate outcomes. However, this aspect seems not to be understood completely yet. It has been claimed that Migdal-Eliashberg theory breaks down at an input coupling $\lambda_0\simeq0.4$, which was concluded by considering a 2D Holstein model\,\cite{Esterlis2018,Chubukov2020}. However, it is by no means clear whether the same limit exists in real materials, and in particular in 3D systems, as was implied, e.g., in Ref.\,\cite{Esterlis2018_2}. Furthermore, other authors have arrived at deviating conclusions: In an early work by Marsiglio \cite{Marsiglio1990} good agreement between QMC and Eliashberg theory was found regardless of coupling strength, provided that the phonon renormalization is taken self-consistently into account. We therefore conclude that the correct picture is currently still elusive, and want to stress again 
that results presented here are solely derived within Migdal's approximation. 

There are several aspects that go beyond the scope of the current work, but are important to be considered in future studies for gaining a better understanding of the formalism. Examples of such extensions would be a non-trivial momentum dependence in the electron-phonon coupling $\lambda_{\mathbf{q}}$, or replacing the Einstein phonon frequency $\Omega_0$ by a wave vector and phonon branch $\nu$ dependent frequency $\omega_{\mathbf{q},\nu}$. Especially a $\nu$-dependence would require a slight generalization of the equations used here. For simplicity, we have neglected any Coulomb repulsion, which should be included in our treatment as well, ideally in a more exploratory manner than just following the most commonly used practice of setting $\mu^{\star}=0.1$\,\cite{Marsiglio2020}. Another major step further would be to extend the non-adiabatic Eliashberg theory of Ref.\,\cite{Schrodi2020_2} by a self-consistent phonon renormalization, which would then have to be done by including up to the second order processes in both electron and phonon self-energies.

\begin{acknowledgments}
	This work has been supported by the Swedish Research Council (VR), the R{\"o}ntgen-{\AA}ngstr{\"o}m Cluster, the Knut and Alice Wallenberg Foundation (grant No.\ 2015.0060), and the Swedish National Infrastructure for Computing (SNIC).	
\end{acknowledgments}

\bibliographystyle{apsrev4-1}

%


\begin{thebibliography}{37}%
	\makeatletter
	\providecommand \@ifxundefined [1]{%
		\@ifx{#1\undefined}
	}%
	\providecommand \@ifnum [1]{%
		\ifnum #1\expandafter \@firstoftwo
		\else \expandafter \@secondoftwo
		\fi
	}%
	\providecommand \@ifx [1]{%
		\ifx #1\expandafter \@firstoftwo
		\else \expandafter \@secondoftwo
		\fi
	}%
	\providecommand \natexlab [1]{#1}%
	\providecommand \enquote  [1]{``#1''}%
	\providecommand \bibnamefont  [1]{#1}%
	\providecommand \bibfnamefont [1]{#1}%
	\providecommand \citenamefont [1]{#1}%
	\providecommand \href@noop [0]{\@secondoftwo}%
	\providecommand \href [0]{\begingroup \@sanitize@url \@href}%
	\providecommand \@href[1]{\@@startlink{#1}\@@href}%
	\providecommand \@@href[1]{\endgroup#1\@@endlink}%
	\providecommand \@sanitize@url [0]{\catcode `\\12\catcode `\$12\catcode
		`\&12\catcode `\#12\catcode `\^12\catcode `\_12\catcode `\%12\relax}%
	\providecommand \@@startlink[1]{}%
	\providecommand \@@endlink[0]{}%
	\providecommand \url  [0]{\begingroup\@sanitize@url \@url }%
	\providecommand \@url [1]{\endgroup\@href {#1}{\urlprefix }}%
	\providecommand \urlprefix  [0]{URL }%
	\providecommand \Eprint [0]{\href }%
	\providecommand \doibase [0]{http://dx.doi.org/}%
	\providecommand \selectlanguage [0]{\@gobble}%
	\providecommand \bibinfo  [0]{\@secondoftwo}%
	\providecommand \bibfield  [0]{\@secondoftwo}%
	\providecommand \translation [1]{[#1]}%
	\providecommand \BibitemOpen [0]{}%
	\providecommand \bibitemStop [0]{}%
	\providecommand \bibitemNoStop [0]{.\EOS\space}%
	\providecommand \EOS [0]{\spacefactor3000\relax}%
	\providecommand \BibitemShut  [1]{\csname bibitem#1\endcsname}%
	\let\auto@bib@innerbib\@empty
	\bibitem [{\citenamefont {Eliashberg}(1960)}]{Eliashberg1960}%
	\BibitemOpen
	\bibfield  {author} {\bibinfo {author} {\bibfnamefont {G.~M.}\ \bibnamefont
			{Eliashberg}},\ }\href@noop {} {\bibfield  {journal} {\bibinfo  {journal}
			{Sov. Phys. JETP}\ }\textbf {\bibinfo {volume} {11}},\ \bibinfo {pages} {696}
		(\bibinfo {year} {1960})}\BibitemShut {NoStop}%
	\bibitem [{\citenamefont {Bardeen}\ \emph {et~al.}(1957)\citenamefont
		{Bardeen}, \citenamefont {Cooper},\ and\ \citenamefont
		{Schrieffer}}]{Bardeen1957}%
	\BibitemOpen
	\bibfield  {author} {\bibinfo {author} {\bibfnamefont {J.}~\bibnamefont
			{Bardeen}}, \bibinfo {author} {\bibfnamefont {L.~N.}\ \bibnamefont {Cooper}},
		\ and\ \bibinfo {author} {\bibfnamefont {J.~R.}\ \bibnamefont {Schrieffer}},\
	}\href {\doibase 10.1103/PhysRev.108.1175} {\bibfield  {journal} {\bibinfo
			{journal} {Phys. Rev.}\ }\textbf {\bibinfo {volume} {108}},\ \bibinfo {pages}
		{1175} (\bibinfo {year} {1957})}\BibitemShut {NoStop}%
	\bibitem [{\citenamefont {Migdal}(1958)}]{Migdal1958}%
	\BibitemOpen
	\bibfield  {author} {\bibinfo {author} {\bibfnamefont {A.~B.}\ \bibnamefont
			{Migdal}},\ }\href@noop {} {\bibfield  {journal} {\bibinfo  {journal} {Sov.
				Phys. JETP}\ }\textbf {\bibinfo {volume} {34}},\ \bibinfo {pages} {996}
		(\bibinfo {year} {1958})}\BibitemShut {NoStop}%
	\bibitem [{\citenamefont {Marsiglio}(1990)}]{Marsiglio1990}%
	\BibitemOpen
	\bibfield  {author} {\bibinfo {author} {\bibfnamefont {F.}~\bibnamefont
			{Marsiglio}},\ }\href {\doibase 10.1103/PhysRevB.42.2416} {\bibfield
		{journal} {\bibinfo  {journal} {Phys. Rev. B}\ }\textbf {\bibinfo {volume}
			{42}},\ \bibinfo {pages} {2416} (\bibinfo {year} {1990})}\BibitemShut
	{NoStop}%
	\bibitem [{\citenamefont {Scalettar}\ \emph {et~al.}(1989)\citenamefont
		{Scalettar}, \citenamefont {Bickers},\ and\ \citenamefont
		{Scalapino}}]{Scalettar1989}%
	\BibitemOpen
	\bibfield  {author} {\bibinfo {author} {\bibfnamefont {R.~T.}\ \bibnamefont
			{Scalettar}}, \bibinfo {author} {\bibfnamefont {N.~E.}\ \bibnamefont
			{Bickers}}, \ and\ \bibinfo {author} {\bibfnamefont {D.~J.}\ \bibnamefont
			{Scalapino}},\ }\href {\doibase 10.1103/PhysRevB.40.197} {\bibfield
		{journal} {\bibinfo  {journal} {Phys. Rev. B}\ }\textbf {\bibinfo {volume}
			{40}},\ \bibinfo {pages} {197} (\bibinfo {year} {1989})}\BibitemShut
	{NoStop}%
	\bibitem [{\citenamefont {Marsiglio}(2020)}]{Marsiglio2020}%
	\BibitemOpen
	\bibfield  {author} {\bibinfo {author} {\bibfnamefont {F.}~\bibnamefont
			{Marsiglio}},\ }\href {\doibase https://doi.org/10.1016/j.aop.2020.168102}
	{\bibfield  {journal} {\bibinfo  {journal} {Ann. Phys.}\ }\textbf {\bibinfo
			{volume} {417}},\ \bibinfo {pages} {168102} (\bibinfo {year}
		{2020})}\BibitemShut {NoStop}%
	\bibitem [{\citenamefont {Chubukov}\ \emph {et~al.}(2020)\citenamefont
		{Chubukov}, \citenamefont {Abanov}, \citenamefont {Esterlis},\ and\
		\citenamefont {Kivelson}}]{Chubukov2020}%
	\BibitemOpen
	\bibfield  {author} {\bibinfo {author} {\bibfnamefont {A.~V.}\ \bibnamefont
			{Chubukov}}, \bibinfo {author} {\bibfnamefont {A.}~\bibnamefont {Abanov}},
		\bibinfo {author} {\bibfnamefont {I.}~\bibnamefont {Esterlis}}, \ and\
		\bibinfo {author} {\bibfnamefont {S.~A.}\ \bibnamefont {Kivelson}},\ }\href
	{\doibase https://doi.org/10.1016/j.aop.2020.168190} {\bibfield  {journal}
		{\bibinfo  {journal} {Ann. Phys.}\ }\textbf {\bibinfo {volume} {417}},\
		\bibinfo {pages} {168190} (\bibinfo {year} {2020})}\BibitemShut {NoStop}%
	\bibitem [{\citenamefont {Esterlis}\ \emph
		{et~al.}(2018{\natexlab{a}})\citenamefont {Esterlis}, \citenamefont
		{Nosarzewski}, \citenamefont {Huang}, \citenamefont {Moritz}, \citenamefont
		{Devereaux}, \citenamefont {Scalapino},\ and\ \citenamefont
		{Kivelson}}]{Esterlis2018}%
	\BibitemOpen
	\bibfield  {author} {\bibinfo {author} {\bibfnamefont {I.}~\bibnamefont
			{Esterlis}}, \bibinfo {author} {\bibfnamefont {B.}~\bibnamefont
			{Nosarzewski}}, \bibinfo {author} {\bibfnamefont {E.~W.}\ \bibnamefont
			{Huang}}, \bibinfo {author} {\bibfnamefont {B.}~\bibnamefont {Moritz}},
		\bibinfo {author} {\bibfnamefont {T.~P.}\ \bibnamefont {Devereaux}}, \bibinfo
		{author} {\bibfnamefont {D.~J.}\ \bibnamefont {Scalapino}}, \ and\ \bibinfo
		{author} {\bibfnamefont {S.~A.}\ \bibnamefont {Kivelson}},\ }\href {\doibase
		10.1103/PhysRevB.97.140501} {\bibfield  {journal} {\bibinfo  {journal} {Phys.
				Rev. B}\ }\textbf {\bibinfo {volume} {97}},\ \bibinfo {pages} {140501}
		(\bibinfo {year} {2018}{\natexlab{a}})}\BibitemShut {NoStop}%
	\bibitem [{\citenamefont {Esterlis}\ \emph
		{et~al.}(2018{\natexlab{b}})\citenamefont {Esterlis}, \citenamefont
		{Kivelson},\ and\ \citenamefont {Scalapino}}]{Esterlis2018_2}%
	\BibitemOpen
	\bibfield  {author} {\bibinfo {author} {\bibfnamefont {I.}~\bibnamefont
			{Esterlis}}, \bibinfo {author} {\bibfnamefont {S.~A.}\ \bibnamefont
			{Kivelson}}, \ and\ \bibinfo {author} {\bibfnamefont {D.~J.}\ \bibnamefont
			{Scalapino}},\ }\href {\doibase 10.1038/s41535-018-0133-0} {\bibfield
		{journal} {\bibinfo  {journal} {npj Quantum Mater.}\ }\textbf {\bibinfo
			{volume} {3}},\ \bibinfo {pages} {59} (\bibinfo {year}
		{2018}{\natexlab{b}})}\BibitemShut {NoStop}%
	\bibitem [{\citenamefont {Bauer}\ \emph {et~al.}(2011)\citenamefont {Bauer},
		\citenamefont {Han},\ and\ \citenamefont {Gunnarsson}}]{Bauer2011}%
	\BibitemOpen
	\bibfield  {author} {\bibinfo {author} {\bibfnamefont {J.}~\bibnamefont
			{Bauer}}, \bibinfo {author} {\bibfnamefont {J.~E.}\ \bibnamefont {Han}}, \
		and\ \bibinfo {author} {\bibfnamefont {O.}~\bibnamefont {Gunnarsson}},\
	}\href {\doibase 10.1103/PhysRevB.84.184531} {\bibfield  {journal} {\bibinfo
			{journal} {Phys. Rev. B}\ }\textbf {\bibinfo {volume} {84}},\ \bibinfo
		{pages} {184531} (\bibinfo {year} {2011})}\BibitemShut {NoStop}%
	\bibitem [{\citenamefont {Alexandrov}(2001)}]{Alexandrov2001}%
	\BibitemOpen
	\bibfield  {author} {\bibinfo {author} {\bibfnamefont {A.~S.}\ \bibnamefont
			{Alexandrov}},\ }\href {\doibase 10.1209/epl/i2001-00492-x} {\bibfield
		{journal} {\bibinfo  {journal} {Europhys. Lett. ({EPL})}\ }\textbf {\bibinfo
			{volume} {56}},\ \bibinfo {pages} {92} (\bibinfo {year} {2001})}\BibitemShut
	{NoStop}%
	\bibitem [{\citenamefont {Meyer}\ \emph {et~al.}(2002)\citenamefont {Meyer},
		\citenamefont {Hewson},\ and\ \citenamefont {Bulla}}]{Meyer2002}%
	\BibitemOpen
	\bibfield  {author} {\bibinfo {author} {\bibfnamefont {D.}~\bibnamefont
			{Meyer}}, \bibinfo {author} {\bibfnamefont {A.~C.}\ \bibnamefont {Hewson}}, \
		and\ \bibinfo {author} {\bibfnamefont {R.}~\bibnamefont {Bulla}},\ }\href
	{\doibase 10.1103/PhysRevLett.89.196401} {\bibfield  {journal} {\bibinfo
			{journal} {Phys. Rev. Lett.}\ }\textbf {\bibinfo {volume} {89}},\ \bibinfo
		{pages} {196401} (\bibinfo {year} {2002})}\BibitemShut {NoStop}%
	\bibitem [{\citenamefont {Perroni}\ \emph {et~al.}(2005)\citenamefont
		{Perroni}, \citenamefont {Cataudella}, \citenamefont {De~Filippis},\ and\
		\citenamefont {Ramaglia}}]{Perroni2005}%
	\BibitemOpen
	\bibfield  {author} {\bibinfo {author} {\bibfnamefont {C.~A.}\ \bibnamefont
			{Perroni}}, \bibinfo {author} {\bibfnamefont {V.}~\bibnamefont {Cataudella}},
		\bibinfo {author} {\bibfnamefont {G.}~\bibnamefont {De~Filippis}}, \ and\
		\bibinfo {author} {\bibfnamefont {V.~M.}\ \bibnamefont {Ramaglia}},\ }\href
	{\doibase 10.1103/PhysRevB.71.054301} {\bibfield  {journal} {\bibinfo
			{journal} {Phys. Rev. B}\ }\textbf {\bibinfo {volume} {71}},\ \bibinfo
		{pages} {054301} (\bibinfo {year} {2005})}\BibitemShut {NoStop}%
	\bibitem [{\citenamefont {Costa}\ \emph {et~al.}(2018)\citenamefont {Costa},
		\citenamefont {Blommel}, \citenamefont {Chiu}, \citenamefont {Batrouni},\
		and\ \citenamefont {Scalettar}}]{Costa2018}%
	\BibitemOpen
	\bibfield  {author} {\bibinfo {author} {\bibfnamefont {N.~C.}\ \bibnamefont
			{Costa}}, \bibinfo {author} {\bibfnamefont {T.}~\bibnamefont {Blommel}},
		\bibinfo {author} {\bibfnamefont {W.-T.}\ \bibnamefont {Chiu}}, \bibinfo
		{author} {\bibfnamefont {G.}~\bibnamefont {Batrouni}}, \ and\ \bibinfo
		{author} {\bibfnamefont {R.~T.}\ \bibnamefont {Scalettar}},\ }\href {\doibase
		10.1103/PhysRevLett.120.187003} {\bibfield  {journal} {\bibinfo  {journal}
			{Phys. Rev. Lett.}\ }\textbf {\bibinfo {volume} {120}},\ \bibinfo {pages}
		{187003} (\bibinfo {year} {2018})}\BibitemShut {NoStop}%
	\bibitem [{\citenamefont {Dee}\ \emph {et~al.}(2019)\citenamefont {Dee},
		\citenamefont {Nakatsukasa}, \citenamefont {Wang},\ and\ \citenamefont
		{Johnston}}]{Dee2019}%
	\BibitemOpen
	\bibfield  {author} {\bibinfo {author} {\bibfnamefont {P.~M.}\ \bibnamefont
			{Dee}}, \bibinfo {author} {\bibfnamefont {K.}~\bibnamefont {Nakatsukasa}},
		\bibinfo {author} {\bibfnamefont {Y.}~\bibnamefont {Wang}}, \ and\ \bibinfo
		{author} {\bibfnamefont {S.}~\bibnamefont {Johnston}},\ }\href {\doibase
		10.1103/PhysRevB.99.024514} {\bibfield  {journal} {\bibinfo  {journal} {Phys.
				Rev. B}\ }\textbf {\bibinfo {volume} {99}},\ \bibinfo {pages} {024514}
		(\bibinfo {year} {2019})}\BibitemShut {NoStop}%
	\bibitem [{\citenamefont {Dee}\ \emph {et~al.}(2020)\citenamefont {Dee},
		\citenamefont {Coulter}, \citenamefont {Kleiner},\ and\ \citenamefont
		{Johnston}}]{Dee2020}%
	\BibitemOpen
	\bibfield  {author} {\bibinfo {author} {\bibfnamefont {P.~M.}\ \bibnamefont
			{Dee}}, \bibinfo {author} {\bibfnamefont {J.}~\bibnamefont {Coulter}},
		\bibinfo {author} {\bibfnamefont {K.~G.}\ \bibnamefont {Kleiner}}, \ and\
		\bibinfo {author} {\bibfnamefont {S.}~\bibnamefont {Johnston}},\ }\href
	{\doibase 10.1038/s42005-020-00413-2} {\bibfield  {journal} {\bibinfo
			{journal} {Commun. Phys.}\ }\textbf {\bibinfo {volume} {3}},\ \bibinfo
		{pages} {145} (\bibinfo {year} {2020})}\BibitemShut {NoStop}%
	\bibitem [{\citenamefont {Kostur}\ and\ \citenamefont
		{Mitrovi{\'{c}}}(1993)}]{Kostur1993}%
	\BibitemOpen
	\bibfield  {author} {\bibinfo {author} {\bibfnamefont {V.~N.}\ \bibnamefont
			{Kostur}}\ and\ \bibinfo {author} {\bibfnamefont {B.}~\bibnamefont
			{Mitrovi{\'{c}}}},\ }\href {\doibase 10.1103/PhysRevB.48.16388} {\bibfield
		{journal} {\bibinfo  {journal} {Phys. Rev. B}\ }\textbf {\bibinfo {volume}
			{48}},\ \bibinfo {pages} {16388} (\bibinfo {year} {1993})}\BibitemShut
	{NoStop}%
	\bibitem [{\citenamefont {Gunnarsson}\ \emph {et~al.}(1994)\citenamefont
		{Gunnarsson}, \citenamefont {Meden},\ and\ \citenamefont
		{Sch\"onhammer}}]{Gunnarsson1994}%
	\BibitemOpen
	\bibfield  {author} {\bibinfo {author} {\bibfnamefont {O.}~\bibnamefont
			{Gunnarsson}}, \bibinfo {author} {\bibfnamefont {V.}~\bibnamefont {Meden}}, \
		and\ \bibinfo {author} {\bibfnamefont {K.}~\bibnamefont {Sch\"onhammer}},\
	}\href {\doibase 10.1103/PhysRevB.50.10462} {\bibfield  {journal} {\bibinfo
			{journal} {Phys. Rev. B}\ }\textbf {\bibinfo {volume} {50}},\ \bibinfo
		{pages} {10462} (\bibinfo {year} {1994})}\BibitemShut {NoStop}%
	\bibitem [{\citenamefont {Nicol}\ and\ \citenamefont
		{Freericks}(1994)}]{Nicol1994}%
	\BibitemOpen
	\bibfield  {author} {\bibinfo {author} {\bibfnamefont {E.~J.}\ \bibnamefont
			{Nicol}}\ and\ \bibinfo {author} {\bibfnamefont {J.~K.}\ \bibnamefont
			{Freericks}},\ }\href {\doibase https://doi.org/10.1016/0921-4534(94)92410-4}
	{\bibfield  {journal} {\bibinfo  {journal} {Physica C: Superconductivity}\
		}\textbf {\bibinfo {volume} {235-240}},\ \bibinfo {pages} {2379 } (\bibinfo
		{year} {1994})}\BibitemShut {NoStop}%
	\bibitem [{\citenamefont {Miller}\ \emph {et~al.}(1998)\citenamefont {Miller},
		\citenamefont {Freericks},\ and\ \citenamefont {Nicol}}]{Miller1998}%
	\BibitemOpen
	\bibfield  {author} {\bibinfo {author} {\bibfnamefont {P.}~\bibnamefont
			{Miller}}, \bibinfo {author} {\bibfnamefont {J.~K.}\ \bibnamefont
			{Freericks}}, \ and\ \bibinfo {author} {\bibfnamefont {E.~J.}\ \bibnamefont
			{Nicol}},\ }\href {\doibase 10.1103/PhysRevB.58.14498} {\bibfield  {journal}
		{\bibinfo  {journal} {Phys. Rev. B}\ }\textbf {\bibinfo {volume} {58}},\
		\bibinfo {pages} {14498} (\bibinfo {year} {1998})}\BibitemShut {NoStop}%
	\bibitem [{\citenamefont {Hague}(2003)}]{Hague2003}%
	\BibitemOpen
	\bibfield  {author} {\bibinfo {author} {\bibfnamefont {J.~P.}\ \bibnamefont
			{Hague}},\ }\href {\doibase 10.1088/0953-8984/15/17/309} {\bibfield
		{journal} {\bibinfo  {journal} {J. Phys.: Condens. Matter}\ }\textbf
		{\bibinfo {volume} {15}},\ \bibinfo {pages} {2535} (\bibinfo {year}
		{2003})}\BibitemShut {NoStop}%
	\bibitem [{\citenamefont {Pietronero}\ and\ \citenamefont
		{Str\"assler}(1992)}]{Pietronero1992_2}%
	\BibitemOpen
	\bibfield  {author} {\bibinfo {author} {\bibfnamefont {L.}~\bibnamefont
			{Pietronero}}\ and\ \bibinfo {author} {\bibfnamefont {S.}~\bibnamefont
			{Str\"assler}},\ }\href {\doibase 10.1209/0295-5075/18/7/010} {\bibfield
		{journal} {\bibinfo  {journal} {Europhys. Lett. ({EPL})}\ }\textbf {\bibinfo
			{volume} {18}},\ \bibinfo {pages} {627} (\bibinfo {year} {1992})}\BibitemShut
	{NoStop}%
	\bibitem [{\citenamefont {Benedetti}\ \emph {et~al.}(1994)\citenamefont
		{Benedetti}, \citenamefont {Grimaldi}, \citenamefont {Pietronero},\ and\
		\citenamefont {Varelogiannis}}]{Benedetti1994}%
	\BibitemOpen
	\bibfield  {author} {\bibinfo {author} {\bibfnamefont {P.}~\bibnamefont
			{Benedetti}}, \bibinfo {author} {\bibfnamefont {C.}~\bibnamefont {Grimaldi}},
		\bibinfo {author} {\bibfnamefont {L.}~\bibnamefont {Pietronero}}, \ and\
		\bibinfo {author} {\bibfnamefont {G.}~\bibnamefont {Varelogiannis}},\ }\href
	{\doibase 10.1209/0295-5075/28/5/010} {\bibfield  {journal} {\bibinfo
			{journal} {Europhys. Lett. ({EPL})}\ }\textbf {\bibinfo {volume} {28}},\
		\bibinfo {pages} {351} (\bibinfo {year} {1994})}\BibitemShut {NoStop}%
	\bibitem [{\citenamefont {Paci}\ \emph {et~al.}(2001)\citenamefont {Paci},
		\citenamefont {Cappelluti}, \citenamefont {Grimaldi},\ and\ \citenamefont
		{Pietronero}}]{Paci2001}%
	\BibitemOpen
	\bibfield  {author} {\bibinfo {author} {\bibfnamefont {P.}~\bibnamefont
			{Paci}}, \bibinfo {author} {\bibfnamefont {E.}~\bibnamefont {Cappelluti}},
		\bibinfo {author} {\bibfnamefont {C.}~\bibnamefont {Grimaldi}}, \ and\
		\bibinfo {author} {\bibfnamefont {L.}~\bibnamefont {Pietronero}},\ }\href
	{\doibase 10.1103/PhysRevB.65.012512} {\bibfield  {journal} {\bibinfo
			{journal} {Phys. Rev. B}\ }\textbf {\bibinfo {volume} {65}},\ \bibinfo
		{pages} {012512} (\bibinfo {year} {2001})}\BibitemShut {NoStop}%
	\bibitem [{\citenamefont {Boeri}\ \emph {et~al.}(2003)\citenamefont {Boeri},
		\citenamefont {Cappelluti}, \citenamefont {Grimaldi},\ and\ \citenamefont
		{Pietronero}}]{Boeri2003}%
	\BibitemOpen
	\bibfield  {author} {\bibinfo {author} {\bibfnamefont {L.}~\bibnamefont
			{Boeri}}, \bibinfo {author} {\bibfnamefont {E.}~\bibnamefont {Cappelluti}},
		\bibinfo {author} {\bibfnamefont {C.}~\bibnamefont {Grimaldi}}, \ and\
		\bibinfo {author} {\bibfnamefont {L.}~\bibnamefont {Pietronero}},\ }\href
	{\doibase 10.1103/PhysRevB.68.214514} {\bibfield  {journal} {\bibinfo
			{journal} {Phys. Rev. B}\ }\textbf {\bibinfo {volume} {68}},\ \bibinfo
		{pages} {214514} (\bibinfo {year} {2003})}\BibitemShut {NoStop}%
	\bibitem [{\citenamefont {Cappelluti}\ \emph {et~al.}(2003)\citenamefont
		{Cappelluti}, \citenamefont {Ciuchi}, \citenamefont {Grimaldi},\ and\
		\citenamefont {Pietronero}}]{Cappelluti2003}%
	\BibitemOpen
	\bibfield  {author} {\bibinfo {author} {\bibfnamefont {E.}~\bibnamefont
			{Cappelluti}}, \bibinfo {author} {\bibfnamefont {S.}~\bibnamefont {Ciuchi}},
		\bibinfo {author} {\bibfnamefont {C.}~\bibnamefont {Grimaldi}}, \ and\
		\bibinfo {author} {\bibfnamefont {L.}~\bibnamefont {Pietronero}},\ }\href
	{\doibase 10.1103/PhysRevB.68.174509} {\bibfield  {journal} {\bibinfo
			{journal} {Phys. Rev. B}\ }\textbf {\bibinfo {volume} {68}},\ \bibinfo
		{pages} {174509} (\bibinfo {year} {2003})}\BibitemShut {NoStop}%
	\bibitem [{\citenamefont {Schrodi}\ \emph
		{et~al.}(2020{\natexlab{a}})\citenamefont {Schrodi}, \citenamefont
		{Oppeneer},\ and\ \citenamefont {Aperis}}]{Schrodi2020_2}%
	\BibitemOpen
	\bibfield  {author} {\bibinfo {author} {\bibfnamefont {F.}~\bibnamefont
			{Schrodi}}, \bibinfo {author} {\bibfnamefont {P.~M.}\ \bibnamefont
			{Oppeneer}}, \ and\ \bibinfo {author} {\bibfnamefont {A.}~\bibnamefont
			{Aperis}},\ }\href {\doibase 10.1103/PhysRevB.102.024503} {\bibfield
		{journal} {\bibinfo  {journal} {Phys. Rev. B}\ }\textbf {\bibinfo {volume}
			{102}},\ \bibinfo {pages} {024503} (\bibinfo {year}
		{2020}{\natexlab{a}})}\BibitemShut {NoStop}%
	\bibitem [{Upp()}]{UppSC}%
	\BibitemOpen
	\href@noop {} {}\bibinfo {note} {The Uppsala Superconductivity (UppSC) code
		provides a package to self-consistently solve the anisotropic, multiband, and
		full-bandwidth Eliashberg equations for frequency-even and odd
		superconductivity mediated by phonons or spin-fluctuations on the basis of
		\textit{ab initio} calculated input.}\BibitemShut {Stop}%
	\bibitem [{\citenamefont {Aperis}\ \emph {et~al.}(2015)\citenamefont {Aperis},
		\citenamefont {Maldonado},\ and\ \citenamefont {Oppeneer}}]{Aperis2015}%
	\BibitemOpen
	\bibfield  {author} {\bibinfo {author} {\bibfnamefont {A.}~\bibnamefont
			{Aperis}}, \bibinfo {author} {\bibfnamefont {P.}~\bibnamefont {Maldonado}}, \
		and\ \bibinfo {author} {\bibfnamefont {P.~M.}\ \bibnamefont {Oppeneer}},\
	}\href {\doibase 10.1103/PhysRevB.92.054516} {\bibfield  {journal} {\bibinfo
			{journal} {Phys. Rev. B}\ }\textbf {\bibinfo {volume} {92}},\ \bibinfo
		{pages} {054516} (\bibinfo {year} {2015})}\BibitemShut {NoStop}%
	\bibitem [{\citenamefont {Bekaert}\ \emph {et~al.}(2018)\citenamefont
		{Bekaert}, \citenamefont {Aperis}, \citenamefont {Partoens}, \citenamefont
		{Oppeneer},\ and\ \citenamefont {Milo\ifmmode \check{s}\else
			\v{s}\fi{}evi\ifmmode~\acute{c}\else \'{c}\fi{}}}]{Bekaert2018}%
	\BibitemOpen
	\bibfield  {author} {\bibinfo {author} {\bibfnamefont {J.}~\bibnamefont
			{Bekaert}}, \bibinfo {author} {\bibfnamefont {A.}~\bibnamefont {Aperis}},
		\bibinfo {author} {\bibfnamefont {B.}~\bibnamefont {Partoens}}, \bibinfo
		{author} {\bibfnamefont {P.~M.}\ \bibnamefont {Oppeneer}}, \ and\ \bibinfo
		{author} {\bibfnamefont {M.~V.}\ \bibnamefont {Milo\ifmmode \check{s}\else
				\v{s}\fi{}evi\ifmmode~\acute{c}\else \'{c}\fi{}}},\ }\href {\doibase
		10.1103/PhysRevB.97.014503} {\bibfield  {journal} {\bibinfo  {journal} {Phys.
				Rev. B}\ }\textbf {\bibinfo {volume} {97}},\ \bibinfo {pages} {014503}
		(\bibinfo {year} {2018})}\BibitemShut {NoStop}%
	\bibitem [{\citenamefont {Schrodi}\ \emph {et~al.}(2019)\citenamefont
		{Schrodi}, \citenamefont {Aperis},\ and\ \citenamefont
		{Oppeneer}}]{Schrodi2019}%
	\BibitemOpen
	\bibfield  {author} {\bibinfo {author} {\bibfnamefont {F.}~\bibnamefont
			{Schrodi}}, \bibinfo {author} {\bibfnamefont {A.}~\bibnamefont {Aperis}}, \
		and\ \bibinfo {author} {\bibfnamefont {P.~M.}\ \bibnamefont {Oppeneer}},\
	}\href {\doibase 10.1103/PhysRevB.99.184508} {\bibfield  {journal} {\bibinfo
			{journal} {Phys. Rev. B}\ }\textbf {\bibinfo {volume} {99}},\ \bibinfo
		{pages} {184508} (\bibinfo {year} {2019})}\BibitemShut {NoStop}%
	\bibitem [{\citenamefont {Schrodi}\ \emph
		{et~al.}(2020{\natexlab{b}})\citenamefont {Schrodi}, \citenamefont {Aperis},\
		and\ \citenamefont {Oppeneer}}]{Schrodi2020_3}%
	\BibitemOpen
	\bibfield  {author} {\bibinfo {author} {\bibfnamefont {F.}~\bibnamefont
			{Schrodi}}, \bibinfo {author} {\bibfnamefont {A.}~\bibnamefont {Aperis}}, \
		and\ \bibinfo {author} {\bibfnamefont {P.~M.}\ \bibnamefont {Oppeneer}},\
	}\href {\doibase 10.1103/PhysRevB.102.014502} {\bibfield  {journal} {\bibinfo
			{journal} {Phys. Rev. B}\ }\textbf {\bibinfo {volume} {102}},\ \bibinfo
		{pages} {014502} (\bibinfo {year} {2020}{\natexlab{b}})}\BibitemShut
	{NoStop}%
	\bibitem [{\citenamefont {Schrodi}\ \emph
		{et~al.}(2020{\natexlab{c}})\citenamefont {Schrodi}, \citenamefont {Aperis},\
		and\ \citenamefont {Oppeneer}}]{SchrodiMultiChan}%
	\BibitemOpen
	\bibfield  {author} {\bibinfo {author} {\bibfnamefont {F.}~\bibnamefont
			{Schrodi}}, \bibinfo {author} {\bibfnamefont {A.}~\bibnamefont {Aperis}}, \
		and\ \bibinfo {author} {\bibfnamefont {P.~M.}\ \bibnamefont {Oppeneer}},\
	}\href@noop {} {} (\bibinfo {year} {2020}{\natexlab{c}}),\ \Eprint
	{http://arxiv.org/abs/2009.08755} {arXiv:2009.08755 [cond-mat.supr-con]}
	\BibitemShut {NoStop}%
	\bibitem [{\citenamefont {Schrodi}\ \emph {et~al.}(2018)\citenamefont
		{Schrodi}, \citenamefont {Aperis},\ and\ \citenamefont
		{Oppeneer}}]{Schrodi2018}%
	\BibitemOpen
	\bibfield  {author} {\bibinfo {author} {\bibfnamefont {F.}~\bibnamefont
			{Schrodi}}, \bibinfo {author} {\bibfnamefont {A.}~\bibnamefont {Aperis}}, \
		and\ \bibinfo {author} {\bibfnamefont {P.~M.}\ \bibnamefont {Oppeneer}},\
	}\href {\doibase 10.1103/PhysRevB.98.094509} {\bibfield  {journal} {\bibinfo
			{journal} {Phys. Rev. B}\ }\textbf {\bibinfo {volume} {98}},\ \bibinfo
		{pages} {094509} (\bibinfo {year} {2018})}\BibitemShut {NoStop}%
	\bibitem [{\citenamefont {Allen}\ and\ \citenamefont
		{Mitrovic}(1983)}]{Allen1983}%
	\BibitemOpen
	\bibfield  {author} {\bibinfo {author} {\bibfnamefont {P.~B.}\ \bibnamefont
			{Allen}}\ and\ \bibinfo {author} {\bibfnamefont {B.}~\bibnamefont
			{Mitrovic}},\ }\enquote {\bibinfo {title} {Theory of superconducting tc},}\ \
	(\bibinfo  {publisher} {Academic Press},\ \bibinfo {year} {1983})\ pp.\
	\bibinfo {pages} {1 -- 92}\BibitemShut {NoStop}%
	\bibitem [{\citenamefont {Moussa}\ and\ \citenamefont
		{Cohen}(2006)}]{Moussa2006}%
	\BibitemOpen
	\bibfield  {author} {\bibinfo {author} {\bibfnamefont {J.~E.}\ \bibnamefont
			{Moussa}}\ and\ \bibinfo {author} {\bibfnamefont {M.~L.}\ \bibnamefont
			{Cohen}},\ }\href {\doibase 10.1103/PhysRevB.74.094520} {\bibfield  {journal}
		{\bibinfo  {journal} {Phys. Rev. B}\ }\textbf {\bibinfo {volume} {74}},\
		\bibinfo {pages} {094520} (\bibinfo {year} {2006})}\BibitemShut {NoStop}%
	\bibitem [{\citenamefont {Veki\ifmmode~\acute{c}\else \'{c}\fi{}}\ \emph
		{et~al.}(1992)\citenamefont {Veki\ifmmode~\acute{c}\else \'{c}\fi{}},
		\citenamefont {Noack},\ and\ \citenamefont {White}}]{Vekic1992}%
	\BibitemOpen
	\bibfield  {author} {\bibinfo {author} {\bibfnamefont {M.}~\bibnamefont
			{Veki\ifmmode~\acute{c}\else \'{c}\fi{}}}, \bibinfo {author} {\bibfnamefont
			{R.~M.}\ \bibnamefont {Noack}}, \ and\ \bibinfo {author} {\bibfnamefont
			{S.~R.}\ \bibnamefont {White}},\ }\href {\doibase 10.1103/PhysRevB.46.271}
	{\bibfield  {journal} {\bibinfo  {journal} {Phys. Rev. B}\ }\textbf {\bibinfo
			{volume} {46}},\ \bibinfo {pages} {271} (\bibinfo {year} {1992})}\BibitemShut
	{NoStop}%
\end{thebibliography}

\appendix

\section{Benchmark calculations}\label{scAppBench}

\begin{figure}[t!]
	\centering
	\includegraphics[width=1\linewidth]{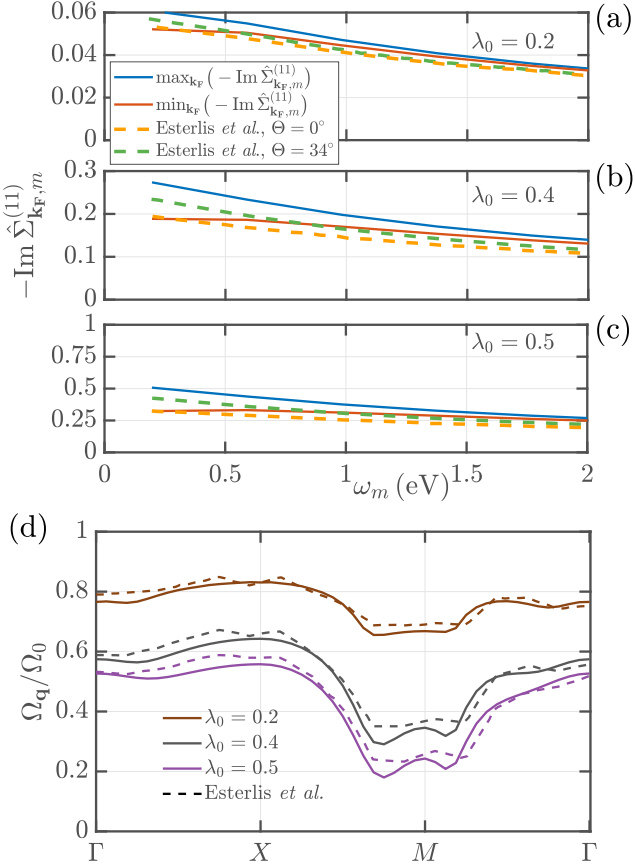}
	\caption{(a-c) Imaginary part of the $(11)$-element of the electron self-energy as function of Matsubara frequencies. Results shown in (a), (b) and (c) are obtained for $\lambda_0=0.2$, $\lambda_0=0.4$ and $\lambda_0=0.5$, respectively. Dashed curves are taken from Ref.\,\cite{Esterlis2018}, at FS angles as written in the legend. Blue and red solid lines correspond to our results computed from Eq.\,(\ref{sigma11}), where we show curves at those FS momenta that produce the maximum (blue) and minimum (red) magnitude for the self-energy. (d) Phonon frequencies, normalized to the bare input $\Omega_0$, along high symmetry lines of the BZ. Different colors represent input couplings as written in the legend. For each $\lambda_0$ we show our results as solid lines, and curves taken from Ref.\,\cite{Esterlis2018} with dashed line style.}\label{benchEsterlis}
\end{figure}
In the following we benchmark our formalism by comparing it to existing literature to show that our implementation is reliable and in agreement with other approaches. Specifically, we compare our results with those of Ref.\,\cite{Esterlis2018}, where the authors compare outcomes of Quantum Monte Carlo calculations with results from Migdal-Eliashberg theory, computing the charge density wave and superconducting susceptibilities. The temperatures considered in this work exceed the phase transition, i.e.\ the system is in the interacting but non-superconducting state. For their Eliashberg calculations the authors include the electron and phonon first-order Feynman diagrams, similar to our present work. The aim in this section is to reproduce Figs.\,3 and 4 published in Ref.\,\cite{Esterlis2018}.

Esterlis {\textit{et al.}\ \cite{Esterlis2018} use a two dimensional tight-binding model with nearest and next-nearest neighbor hopping energies $t$ and $t^{\prime}$. Their ratio is taken as $t^{\prime}/t=-0.3$. The electron filling is fixed at $n=0.8$. For both Figures that we are interested in, the temperature is $T=t/16$, the adiabaticity ratio is $\Omega_0/E_F=0.1$ (with $E_F$ the Fermi energy) and the coupling strength is chosen from $\lambda_0\in\{0.2,0.4,0.5\}$. The functional form of their tight-binding approach reads
\begin{align}
\xi_{\mathbf{k}}= -2t\big[\cos(k_x)+\cos(k_y)\big] - 4t^{\prime} \cos(k_x)\cos(k_y) - \mu
\end{align}
for the electron energies\,\cite{Vekic1992}. The chemical potential $\mu$ has to be adjusted so as to fix the electron filling at the chosen value. In contrast to our theory presented here, Esterlis {\em et al.}\ \cite{Esterlis2018} have focused on properties above $T_c$ using a scalar function $\Sigma(\mathbf{k},\omega_m)$, while our Nambu formalism allows us to directly access superconducting properties. We can make contact between the two formulations by considering only the $(11)$ matrix element of $\hat{\Sigma}_{\mathbf{k},m}$,
\begin{align}
-\mathrm{Im}\hat{\Sigma}_{\mathbf{k},m}^{(11)} = \omega_m (Z_{\mathbf{k},m}-1) ~,\label{sigma11}
\end{align}
which is equivalent to the scalar self-energy of Ref.\,\cite{Esterlis2018}.

In Fig.\,\ref{benchEsterlis}(a-c) we show our results for Eq.\,(\ref{sigma11}) as solid lines in all three panels, corresponding to different choices of $\lambda_0$. We plot $-\mathrm{Im}\hat{\Sigma}_{\mathbf{k},m}^{(11)}$ at those FS momenta, that produce the maximal (blue curve) and minimal (red curve) magnitude of the result. As comparison we extracted results from Ref.\,\cite{Esterlis2018} at two different FS angles, see the orange and green dashed lines. It is directly apparent that the curves coincide to a very good degree.

Next we attempt to reproduce the renormalized phonon frequency spectrum of Ref.\,\cite{Esterlis2018}, which can be obtained with the same set of parameters as described above. We show our results for $\lambda_0=0.2$, $\lambda_0=0.4$, and $\lambda_0=0.5$ in Fig.\,\ref{benchEsterlis}(d) via solid brown, gray and purple lines, respectively. As before, we find  good agreement with results by Esterlis {\em et al.}, plotted as dashed curves in similar color code.

\end{document}